\documentclass[12pt]{article}
\pdfoutput=1
\usepackage{jheppub}
\usepackage{amssymb,amsmath,amstext,amsfonts}
\usepackage{bbold,ulem,bm}
\usepackage{graphics}
\usepackage{mathtools}
\usepackage{hyperref}
\usepackage{color}
\usepackage{ulem}

\usepackage{graphicx}
\usepackage{mathrsfs}
\usepackage{ytableau}

\newcommand{\beq}{\begin{equation}}
\newcommand{\eeq}{\end{equation}}
\newcommand{\bal}{\begin{aligned}}
\newcommand{\eal}{\end{aligned}}
\newcommand{\Lag}{{\mathcal{L}}}

\newcommand{\D}{{\mathcal{D}}}

\newcommand{\td}{{\mathrm{t.d.}}}

\renewcommand{\arraystretch}{2}

\definecolor{rougef}{rgb}{0.99,0,0}
\definecolor{vertf}{rgb}{0,0.8,0.2}
\definecolor{bleuf}{rgb}{0,0,0.8}

\begin{document}

\title{Consistent deformations of free massive field theories in the Stueckelberg formulation} 

\author[a]{Nicolas Boulanger,}
\affiliation[a]{Groupe de M\'ecanique et Gravitation, Physique th\'eorique 
et math\'ematique, Universit\'e de Mons -- UMONS, 20 place du Parc, 7000 Mons,
Belgique}
\author[b,c]{C\'edric Deffayet,}
\affiliation[b]{Sorbonne Universit\'es, UPMC Univ.\ Paris 6 and CNRS, UMR 7095, Institut d'Astrophysique de Paris, GReCO, 
98bis boulevard Arago, 75014 Paris, France}
\affiliation[c]{IHES, Le Bois-Marie, 35 route de Chartres, 91440 Bures-sur-Yvette, France}
\author[b]{Sebastian Garcia-Saenz,}
\author[a]{Lucas Traina}

\abstract{Cohomological techniques within the 
Batalin--Vilkovisky (BV) extension of the 
Becchi--Rouet--Stora--Tyutin (BRST) formalism 
have proved invaluable for classifying consistent deformations of gauge theories.
In this work we investigate the application of this idea to massive field theories 
in the Stueckelberg formulation. Starting with a collection of free massive vectors, 
we show that the cohomological method reproduces the cubic and quartic 
vertices of massive Yang--Mills theory. In the same way, taking a Fierz--Pauli 
graviton on 
a maximally symmetric space as the starting point, we are able to recover the 
consistent cubic vertices of nonlinear massive gravity. The formalism further 
sheds light on the characterization of Stueckelberg gauge theories, 
by demonstrating for instance that the gauge algebra of such models 
is necessarily Abelian and that they admit a Born--Infeld-like formulation 
in which the action is simply a combination of the gauge-invariant structures
of the free theory.
}


\maketitle


\section{Introduction} \label{sec:intro}

Given the free theory of some set of fields, what are the possible interactions 
that one can add in a consistent manner? This question was first addressed decades ago in the context of Einstein gravity, 
with several works --- see e.g.\ \cite{Gupta:1954zz,Weinberg:1965rz,Deser:1969wk} 
and the references given by Preskill and Thorne 
in their foreword to the Feynman lectures on Gravitation \cite{FeynmanLecturesGravitation} --- 
showing via different methods that, under certain minimal 
assumptions, general relativity (GR) can be derived as the unique 
nonlinear extension of the Fierz--Pauli action for a massless spin-2 particle. 
A systematic analysis of the problem of introducing consistent interactions 
in a gauge theory was given in \cite{Berends:1984rq}.  
Perhaps the most important landmark in this program was the 
cohomological reformulation \cite{Barnich:1993vg} of the 
analysis \cite{Berends:1984rq}  
within the Batalin--Vilkovisky (BV) antifield formalism 
\cite{Batalin:1981jr,Batalin:1984jr}.
The latter approach extends the Becchi--Rouet--Stora--Tyutin (BRST)
formalism \cite{Becchi:1975nq,Tyutin:1975qk}
by the adjunction of extra structures and antifields. 
In particular, the cohomological approach \cite{Barnich:1993vg} generalizes  
the approach of Wald, who studied interactions for massless spin-1 and spin-2 fields 
by demanding consistency of the gauge algebra \cite{Wald:1986bj,Cutler:1986dv} (see also \cite{Deser:1963zzc} for an earlier treatment of spin-1 fields). 
The cohomological antifield method goes one step further as it unifies 
into a purely algebraic framework the problems of deforming consistently 
both the gauge symmetry and the action functional of a theory.

Originally introduced as an instrument to examine quantum aspects of 
gauge theories with an open gauge
algebra \cite{deWit:1978hyh,Batalin:1981jr,Batalin:1984jr} 
such as their renormalizability and anomalies, the BRST-BV formalism
has proved extremely powerful also at the classical 
level, see \cite{Barnich:2000zw} for a review. 
In this setting, full classifications of consistent interaction vertices have 
been achieved in a number of theories including Yang--Mills 
\cite{Barnich:1994db,Barnich:1994mt}, massless vector-scalar 
models \cite{Barnich:2017nty,Barnich:2018nqa}, 
Einstein and Weyl multi-gravity 
\cite{Boulanger:2000rq,Boulanger:2001he} 
as well as pure supergravity \cite{Boulanger:2018fei}. 
See \cite{Henneaux:2012wg,Henneaux:2013gba,Bizdadea:2015yip,Bizdadea:2016hze} 
for other references where the cohomological approach for consistent 
deformations of classical actions was used.
A common property of all these examples is the presence of gauge symmetries. 
This would seem to preclude its use in the study of {\it massive} theories, 
which of course do not possess any gauge symmetry in their minimal 
covariant formulation.
It is, however, well understood that gauge invariance does not have any 
true physical meaning, but rather encodes a redundancy in the phase space 
of a theory. For massless and partially massless field theories, 
such a redundancy is useful because it allows for a description that is 
manifestly local and Lorentz invariant --- or invariant under possibly other spacetime isometries. This is not a problem in 
massive theories, 
but even in this case one may wish to work with a gauge invariant 
formulation as a tool for studying some of their physical properties. 
This method is known as the Stueckelberg procedure. It simply amounts to 
introducing an additional set of fields and corresponding gauge symmetries 
in such a way that the propagating degrees of freedom remain unchanged.
Many successful instances where this approach was followed to clarify 
and discover properties of massive systems can be found 
in, e.g., 
\cite{Gherghetta:2002nr,ArkaniHamed:2002sp,Zinoviev:2006im,Quadri:2010uk,deRham:2010ik,
deRham:2010kj,Hassan:2011vm,deRham:2011rn,Rahman:2011ik,Hassan:2012qv,
Gabadadze:2013ria,Alberte:2013sma,Noller:2013yja,Gao:2014ula,
Buchbinder:2015mta,Hinterbichler:2015soa,Zinoviev:2016mxh,Zinoviev:2018juc}.
There is therefore no fundamental limitation to the application of the 
cohomological method of \cite{Barnich:1993vg} to examine consistent 
deformations of massive field theories, and it is the purpose of the present 
paper to initiate a systematic implementation of this idea.
For completeness, we mention that another algebraic method for the study 
of consistent deformations of massive and massless field theories 
was provided in \cite{Kaparulin:2012px} and used, e.g., 
in \cite{Cortese:2013lda,Cortese:2017ieu}. We note that the method exposed 
in \cite{Kaparulin:2012px} is more general, in the sense that it 
does not require any Lagrangian field equations as a starting point. 

Concretely, our main goals will be to understand the general properties that 
characterize Stueckelberg gauge theories in the BRST-BV language 
and to see whether the cohomological approach of \cite{Barnich:1993vg} 
can indeed be successfully applied in classifying interaction 
vertices. We address the first point in section \ref{sec:massivebv}, 
where in addition to an elementary review of the formalism we draw 
some general 
conclusions that are valid for any theory with Stueckelberg fields. 
We then 
proceed to illustrate our techniques by examining two relatively simple yet 
interesting models: massive Yang--Mills theory (section \ref{sec:massiveym}) 
and massive gravity (section \ref{sec:massivegrav}).

An important conclusion of our analysis is that the cohomological 
formalism of \cite{Barnich:1993vg} 
allows for a precise characterization of the gauge structure 
of massive theories. We will show in particular that Stueckelberg 
models possess the following properties:
\begin{itemize}
\item They retain the Abelian nature of the initial gauge algebra;
\item Their actions are of the Abelian Born--Infeld type for an appropriate 
choice of field variables, i.e., are expressible 
solely in terms of gauge invariant building blocks of the Abelian,  
free gauge theory;
\item The gauge transformation laws of the Stueckelberg fields can 
always be reduced to a pure shift, with no field-dependent corrections;
\item After an appropriate decoupling limit, 
our procedure gives a sum of two Lagrangians, 
one of which gives a nonlinear sigma model, i.e., a 
non-linear realisation of some non-Abelian group.
\end{itemize}
It is worth emphasizing that the previous properties are often not manifest, 
and for this reason they are perhaps rather unexpected. Indeed, Stueckelberg 
theories are usually formulated in a way that makes the study of their degrees 
of freedom more transparent, but with the fields transforming non-trivially under 
the gauge symmetries. Our results show, however, that it is always possible to 
perform a redefinition of the fields and gauge parameters so that the 
transformation laws of the fields and the invariances of the action 
greatly simplify. 
We are thus uncovering here what may be thought of as a dual picture 
in which the gauge structure of the theory becomes very clear --- in fact 
Abelian --- but with the catch being that the vertices often involve more 
derivatives than in the standard parametrization, 
thereby making power counting and the identification of the relevant 
interaction scales more subtle.

As advertised, in order to make these notions more tangible and to test the 
usefulness of the BRST-BV deformation procedure in the context of massive field 
theories, we consider massive Yang--Mills theory and massive gravity as our 
first case studies. For the first model the starting point is the free action 
for a collection of vector fields $A_{\mu}^a$ and as many Stueckelberg 
scalars $\pi^a\,$. We push the calculation up to second order in the 
deformation analysis, allowing us to recover the known cubic and quartic 
vertices of the gauge invariant formulation of massive Yang--Mills theory, 
derived for instance in \cite{Zinoviev:2006im}, which we also confirm by 
comparing with the full result obtained by starting with the interacting 
theory and performing a non-linear Stueckelberg replacement. 
Lastly we make explicit the field redefinition that takes the ``standard'' 
action into its Born--Infeld form, where the vertices are written as 
contraction of tensors that are invariant under the gauge symmetries 
of the {\it free} theory.

We next repeat the analysis for the case of a massive spin-2 field $h_{\mu\nu}$ on a maximally symmetric space. The calculations here are more involved and we only study the first order deformations that lead to the 3-point vertices of the theory. Here too we are able to rederive all the known structures (with up to four derivatives) previously classified for instance in \cite{Hinterbichler:2017qyt}, and again make explicit the Born--Infeld formulation of the action and the field redefinition connecting it to the standard parametrization. We also briefly examine the special case where the graviton mass is tuned relative to the cosmological constant as 
$m^2=\Lambda/(D-1)$ (with $D$ the spacetime dimension),
for which the spin-2 particle becomes {\it partially massless} \cite{Deser:1983mm,Deser:2001pe,Deser:2001us,Gabadadze:2008uc}. 
We highlight the peculiarities of the partially massless theory relative to the generic massive set-up, and as a by-product we recover the unique cubic vertex that is known to exist only in four dimensions (see e.g.\ \cite{deRham:2013wv}).


\section{Massive theories in the BRST-BV formalism} \label{sec:massivebv}

\subsection{BRST-BV deformation method}

The cohomological reformulation of the deformation procedure exposed 
in \cite{Berends:1984rq} was proposed in \cite{Barnich:1993vg}. 
It exploits the antifield formalism 
\cite{Batalin:1981jr,Batalin:1984jr} for gauge theories.
A detailed and accessible introduction to the BRST-BV deformation procedure  
can be found in \cite{Henneaux:1997bm}; see also the introduction 
of \cite{Boulanger:2000rq}. 
Here we reproduce the essential 
aspects in order to fix the necessary notation and describe the main 
steps of our approach in a self-contained manner as possible. 

Consider a theory for a set of gauge fields $\varphi^i$ defined by an action 
$S[\varphi^i]$ that is invariant under the infinitesimal gauge symmetries
\beq \label{eq:bv gaugesym}
\delta_{\epsilon}\varphi^i=R^i{}_{\alpha}\,\epsilon^{\alpha}\;,
\eeq
where, using De Witt's notation, a summation over repeated indices also means 
that an integral over an omitted dummy coordinate is implied, 
so that \eqref{eq:bv gaugesym} 
gives a local relation depending on the gauge parameters $\epsilon^{\alpha}$ 
and their derivatives up to some finite order. The operator $R^i{}_{\alpha}$ may 
depend on the gauge fields and some of their successive derivatives. 

In the BRST formalism \cite{Becchi:1975nq,Tyutin:1975qk}, 
one proceeds by associating a ghost field $C^\alpha$ to every gauge parameter 
$\epsilon^\alpha\,$, with a shift in the Grassmann parity: 
$|C^\alpha| = |\epsilon^\alpha| +1$ (mod 2).
If the theory is reducible, such as for $p\,$-form gauge theories with $p>1\,$, 
one also introduces
a hierarchy of higher-degree ghosts (ghosts of ghosts). We will not discuss 
these cases here and stick to irreducible gauge theories. 
In the antifield extension of the BRST formalism due to  Batalin and Vilkovisky, 
to the gauge fields $\varphi^i$ and the BRST ghosts $C^\alpha$ one  
associates the antifields $\varphi^{*}_i$ and $C^{*}_\alpha\,$, respectively.\footnote{The antifields $C^{*}_\alpha$ are often 
called 
{\it{antighosts}}, in this context, although they should not be confused 
with the antighosts that appear in the Faddeev--Popov procedure. 
The latter form trivial pairs with 
the Lautrup--Nakanishi fields and can be eliminated from the BRST cohomology, 
see e.g.\ \cite{Henneaux:1992ig,Weinberg:1996kr}.} 
For concreteness we will take both the fields $\varphi^i$ and the 
gauge parameters $\epsilon^{\alpha}$ to be bosonic, thereby 
excluding the discussion of supergravity theories, although the general 
case can be treated in much the same way modulo some obvious changes, 
see for example the pedagogical review \cite{Gomis:1994he}. 
In this situation the ghost antifields are Grassmann-even or 
commuting variables as well, while the ghosts $C^\alpha$ and 
antifields $\varphi^*_i$ are Grassmann-odd or anticommuting. 
Next we introduce two gradings, the ghost number ``${\rm gh}$'' 
and the antifield number ``${\rm antifld}$'', according to 
table \ref{tab:intro gradings}. It is also useful to keep track 
of the number of differentiated or
undifferentiated ghosts (not counting the associated antifields), 
via the pureghost number
``${\rm puregh}$''. We collectively denote the fields and 
ghosts by the variables $\Phi^A\,$, while their associated antifields 
are denoted by $\Phi^*_A\,$. 

\setlength{\tabcolsep}{24pt}
\renewcommand{\arraystretch}{1.5}
\begin{table}
\centering
\begin{tabular}{l c c c}
\hline
  & ${\rm gh}$ & ${\rm antifld}$ & ${\rm puregh}$\\
\hline\hline
Fields $\varphi^i$ & $0$ & $0$ & $0$\\
Ghosts $C^\alpha$ & $1$ & $0$ & $1$\\
Antifields $\varphi^{*}_i$ & $-1$ & $1$ & $0$\\
Ghost antifields $C^{*}_\alpha$ & $-2$ & $2$ & $0$\\
\hline
\end{tabular}
\label{tab:intro gradings}
\caption{Ghost, antifield and pureghost quantum numbers.}
\end{table}

To the initial theory with gauge-invariant action $S[\varphi^i]\,$, 
one associates a BV functional $W[\Phi^A,\Phi^*_A]\,$ in the following way:
\beq \label{eq:bv bvfunc}
W[\Phi^A,\Phi^*_A]=S[\varphi^i]+\varphi^{*}_i\,R^i{}_{\alpha}\,C^{\alpha}
+\tfrac{1}{2}\,C^{*}_{\gamma}\,f^{\gamma}_{\phantom{\gamma}\alpha\beta}\,
C^{\alpha}C^{\beta} 
+ \tfrac{1}{4}\,\varphi^{*}_i\varphi^{*}_j\,M^{ij}_{\alpha\beta}\,C^{\alpha}C^{\beta}+\cdots\;.
\eeq
By construction, as it is required to start with the classical action and 
to have definite ghost number and Grassmann parity, 
the BV functional $W$ is defined to be bosonic (Grassmann-even) 
with ghost number zero. More importantly, it is required to satisfy 
what is called the {\it classical master equation}:
\beq \label{eq:bv mastereq}
(W,W)=0\,,
\eeq
where the antibracket, also called BV bracket, 
is defined by\footnote{Right and left functional derivatives are defined via
\beq
\delta A=\int\frac{\delta^RA}{\delta\Phi^A}\,\delta\Phi^A +
\int \frac{\delta^RA}{\delta\Phi^*_A}\,\delta\Phi^*_A =  
\int \delta\Phi^A\,\frac{\delta^LA}{\delta\Phi^A}+
\int\delta\Phi^*_A\,\frac{\delta^LA}{\delta\Phi^*_A}\;.\nonumber
\eeq
The distinction between left and right derivatives of course only makes a difference when the derivative is with respect to a 
fermionic 
(Grassmann-odd) variable.}
\beq
(A,B)=\frac{\delta^RA}{\delta\Phi^A}\,\frac{\delta^LB}{\delta\Phi^{*}_A}
-\frac{\delta^RA}{\delta\Phi^{*}_A}\,\frac{\delta^LB}{\delta\Phi^A}\;.
\eeq
At zero antifield number, the master equation yields the Noether identities 
associated with the original gauge symmetries of the action,
\beq
\frac{\delta S}{\delta \varphi^i}\,R^i{}_{\alpha}=0\;.
\eeq
Next, the terms with antifield number one in \eqref{eq:bv mastereq} produce
\beq
R^j{}_{\alpha}\,\frac{\delta R^i{}_{\beta}}{\delta \varphi^j}-R^j{}_{\beta}\,\frac{\delta R^i{}_{\alpha}}{\delta \varphi^j}
=R^i{}_{\gamma}\,f^{\gamma}_{\phantom{\gamma}\alpha\beta}
+ \frac{\delta S}{\delta \varphi^j}\,M^{ji}_{\alpha\beta}\;,
\eeq
which is nothing but the gauge algebra of the transformations 
\eqref{eq:bv gaugesym}, which gives the interpretation of $f^{\gamma}_{\phantom{\gamma}\alpha\beta}$ 
as ``structure functionals'', since in general they are operators that may 
depend on the fields. 
The presence of the term proportional to $M^{ij}_{\alpha\beta}$ implies 
that the algebra will in general be ``open'', meaning that two gauge 
transformations will only close upon use of the equations of motion. 
Similarly, at the following order one finds the Jacobi identity satisfied 
by the structure functionals, and continuing in this manner generates a 
tower of consistency relations involving the higher order tensors 
--- the ellipsis in \eqref{eq:bv bvfunc} --- that characterize the gauge 
group of the theory.
We refer to \cite{Gomis:1994he} for detailed discussions and review. 

The main idea of the deformation analysis is to revert this story. 
The full action $S$ and its gauge symmetries are unknown, 
and we seek to determine them by perturbatively solving the master equation, 
knowing an initial action $S_0$ invariant under the gauge transformations 
\begin{equation}
\label{freegaugetransfos}
\delta_0\varphi^i = R_0{}^i{}_{\alpha}\,\epsilon^\alpha\;.    
\end{equation}
The BV functional, as we recalled, encodes all the information about the gauge 
structure of the theory. Hence the BV formalism is equivalent to other more 
direct approaches that aim at determining a theory based on the action and 
gauge transformation, although we will see that it is in many ways more 
powerful.

In order to solve the classical master equation, we consider the 
functional $W$ as a perturbation series in some overall deformation 
parameter $g\,$, i.e., 
\beq \label{eq:bv wseries}
W=W_0+g\,W_1+g^2\,W_2+\cdots\;.
\eeq
Here $W_0$ corresponds to the BV functional associated with the theory 
that is already known and that one wishes to deform perturbatively. 
In the scenarios we focus on in this paper, the known theories will 
be free, so that the deformation procedure amounts to studying the 
consistent interaction vertices one can add to it. 
However, the formalism is not restricted to this case, 
since for instance one can also apply it to known models which are 
themselves already interacting, see for example 
\cite{Barnich:2017nty,Barnich:2018nqa} for recent analyses.

\subsection{Local BRST cohomology}

The master equation approach becomes particularly powerful when rephrased 
as a cohomological problem \cite{Barnich:1993vg}. 
The BV functional $W_0$ for the free theory $S_0$ is viewed as 
the generator of BRST transformations, in the sense that
\beq
s{A} \coloneqq (W_0, A)\;,
\eeq
for any local functional ${ A}\,$, with $s$ denoting the BRST 
differential of the free theory.
By a local functional $A\,$, one means the integral 
$\int a$ of a $D\,$-form $a$ that depends on the fields (including the 
ghosts), their associated antifields and their derivatives up to some 
arbitrary but finite order, which we indicate by the notation 
$a = a([\Phi],[\Phi^*],dx)\,$. 
Moreover, in this work we assume that all the fields and their 
derivative vanish at infinity, or alternatively that they have 
compact support, which enables us to discard all boundary terms. 
With this assumption, any local $D\,$-form $a([\Phi],[\Phi^*],dx)$
is equivalent to $a([\Phi],[\Phi^*],dx) + d b([\Phi],[\Phi^*],dx)$
where $b([\Phi],[\Phi^*],dx)$ is a local $(D-1)\,$-form and where
$d$ is the total exterior differential
$d=dx^\mu\partial_\mu\,$, where
$\partial_\mu = \frac{\partial\phantom{x^\mu}}{\partial x^\mu}
+\partial_\mu z^M \frac{\partial\phantom{z^M}}{\partial z^M}+\dots$ is the
total derivative that takes into account the spacetime dependence of
the fields and antifields that we have collectively denoted by
$z^M\coloneqq(\Phi^A,\Phi^*_A)\,$.  We use conventions whereby 
$s$ anticommutes with $d\,$, or equivalently, for the type of 
theories we will deal with in this paper,  
$dx^\mu C^\alpha +  C^\alpha dx^\mu = 0 $ and 
$dx^\mu \varphi^*_i +  \varphi^*_i dx^\mu = 0 \,$.
One has the following isomorphism of cohomological classes,
$H^g(s)\cong H^{g,D}(s|d)\,$, where $H^g(s)$ denotes the
cohomology of $s$ in the class of ghost number $g$ local functionals
and $H^{g,D}(s|d)$ is the cohomology of the differential $s\,$, 
at ghost number $g\,$, in the space of local $D\,$-forms. 
In other words, the cohomology class in $H^{g,D}(s|d)$ 
is defined up to the $\sim$ equivalence 
relation by a representative solution $a^{g,D}$ of
\begin{equation}
  \label{eq:9a}
  s a^{g,D} +d a^{g+1,D-1}=0,\quad a^{g,D}\sim a^{g,D}+s b^{g-1,D}
  +db^{g,D-1}. 
\end{equation}
The first superscript refers to the ghost number and the second one 
to the form degree.

The basic properties of the BRST differential $s$ then follow from the 
properties of the antibracket; in particular the nilpotency, $s^2=0\,$, 
is a consequence of the (graded) Jacobi identity and of the master 
equation $(W_0,W_0)=0\,$.

Given a free theory with action $S_0[\varphi^i]$ invariant under 
irreducible gauge symmetries as in \eqref{freegaugetransfos},
we can write the effect of a BRST transformation on the fields by 
decomposing $s=\gamma+\delta\,$, such that
\beq
\gamma\varphi^i=R_0{}^i{}_{\alpha}\,C^{\alpha}\,,\qquad 
\delta\varphi^{*}_i=\frac{\delta S_0}{\delta\varphi^i}\,,\qquad 
\delta C^{*}_\alpha=\varphi^*_i\,R_0{}^i{}_{\alpha}\;,
\eeq
while the action of $\gamma$ and $\delta$ on the variable not shown is by definition zero. It follows immediately that
\beq
\gamma^2=0\,,\qquad \delta^2=0\,,\qquad \gamma\delta+\delta\gamma=0\,,
\eeq
which together again amount to the nilpotency of $s\,$.
We recall that, with our conventions, $\{\gamma , d\} = 0 = \{\delta , d\}\,$.
Observe that $\delta$ decreases the antifield number by one unit, 
while $\gamma$ does not change it.
From eqs.\ \eqref{eq:bv mastereq} and \eqref{eq:bv wseries}, 
one has that the master equation, up to order $g^2\,$, yields
\beq
sW_0=0\,,\qquad sW_1=0\,,\qquad sW_2=-\tfrac{1}{2}(W_1,W_1)\;.
\eeq
The first of this holds trivially of course since the free action $S_0$ 
and its gauge invariance is known. The second equation then states that the 
first order deformation of the BV functional is BRST-closed. 
Any BRST-exact expression $W_1=s B$ for a local 
functional $B$ at ghost 
number $-1$ is of course a solution, but it is trivial in that it can 
be obtained from $W_0$ by means of a generalized non-linear field redefinition, 
where by ``generalized'' we mean one that may involve the ghosts and 
antifields as well. At the level of the original action such transformations 
translate into redefinitions of the field variables and/or gauge parameters. 
This leads to the conclusion that 
non-trivial first order (in the deformation parameter $g$) 
solutions of the master 
equation belong to the local cohomological group of $s$ at ghost number zero, 
denoted by $H^{0,D}(s\vert d)\,$. 
The same considerations apply to all higher order deformations. 
For instance $W_2$ now satisfies an inhomogeneous equation, but once a 
particular solution is found the homogeneous part will admit trivial 
terms arising from field redefinitions of $W_2\,$.

\subsection{Stueckelberg procedure} \label{subsec:stueck}

\paragraph{Generalities.}
The Stueckelberg procedure has proved to be a convenient technique to understand a number of important aspects relevant to 
the consistency of massive field theories, for example in counting the degrees of freedom, in analyzing their stability, and in 
obtaining certain high-energy regimes (the so-called decoupling limits). More relevant to us is its application to the analysis of 
deformations of free theories \cite{Zinoviev:2006im}, although an implementation of this idea in the context of the BRST-BV formalism has not been worked out yet, a gap that we want to fill with the present work.

A pedagogical review of the Stueckelberg method through several examples is given in \cite{Hinterbichler:2011tt}; see also the references given in the introduction for other applications. Here we restrict ourselves with a brief abstract explanation, 
which will be helpful in proving certain general attributes that characterize Stueckelberg gauge theories. Take again an action 
$S[\varphi]$ of the generic form
\beq
S=\int d^Dx\,\left(\Lag_k+\Lag_m\right)\,,
\eeq
where $\Lag_k$ is invariant under the gauge transformation
\beq
\varphi^i\to\varphi^{i\,\prime}=F^i([\varphi],[\epsilon])\,,
\eeq
with local parameters $\epsilon^{\alpha}$ and such that ${F}^i([\varphi],0)=\varphi^i\,$. The second term, $\Lag_m$, however 
breaks this symmetry explicitly. (The subscripts stand for ``kinetic'' and ``mass'', as this is the typical situation, although here 
we do not make any assumptions regarding the dimensions of the operators entering in $\Lag_k$ and $\Lag_m$.) We can 
restore the invariance of the action by introducing a set of fields $\sigma^{\alpha}$, the ``Stueckelberg fields'', by performing 
the replacement
\beq
\varphi^i\to{F}^i([\varphi],[\sigma])\,,
\eeq
everywhere in $S[\varphi]\,$. 
The kinetic term is of course unaffected, while the mass term changes as
\beq
\Lag_m(\varphi)\to \Lag_m'([\varphi],[\sigma])
\coloneqq\Lag_m \left({F}([\varphi],[\sigma])\right)\,.
\eeq
The claim is that the new action has a gauge symmetry, 
written infinitesimally as
\beq
\delta_{\epsilon}\varphi^i = R^i{}_{\alpha}\epsilon^\alpha\,,\qquad \delta_{\epsilon}\sigma^{\alpha} = 
S^\alpha_{\beta}([\varphi],
[\sigma])\epsilon^\beta\,,
\eeq
with 
$R^i{}_{\alpha}\coloneqq\delta{F}^i/\delta\sigma^{\alpha}\big|_{\sigma=0}\,$. 
Without further knowledge of the gauge 
group we cannot say much about the form of $S^\alpha_{\beta}\,$, 
but a crucial property is the fact that
\beq \label{eq:zerothS}
S^\alpha_{\beta}([\varphi],0)=-\delta^\alpha_\beta\,.
\eeq
To see this, observe that the condition of invariance of the Lagrangian $\Lag_m'$ implies the Noether identity
\beq \label{eq:stueck noether}
\frac{\delta{F}^i}{\delta\varphi^j}\,R^j{}_{\alpha}+\frac{\delta{F}^i}{\delta\sigma^\beta}\,S^\beta_\alpha=0\,,
\eeq
since $\Lag_k$ is invariant by itself. Eq.\ \eqref{eq:zerothS} then follows by evaluating at $\sigma^\alpha=0$. Thus, 
perturbatively we have that
\beq \label{eq:stueck sigma}
\delta_{\epsilon}\sigma^{\alpha}=-\epsilon^\alpha+\cdots\,,
\eeq
for any Stueckelberg gauge theory. This simple result already gives us quite a lot of information.\footnote{Eq.\ \eqref{eq:stueck 
sigma} can also be understood from the fact that Stueckelberg fields can be interpreted as the Goldstone modes associated with 
the breaking of the gauge group down to its global subgroup \cite{Goon:2014ika,Goon:2014paa}.}
It tells us that there exists a gauge condition, the so-called unitary gauge, in which $\sigma^\alpha=0$, which establishes 
that the theory is dynamically equivalent to the one we started with. Moreover, unitary gauge can be applied directly at the level 
of the action (it is a ``good'' gauge choice in the terminology of \cite{Lagos:2013aua}), since the equations of motion of the Stueckelberg fields are always implied by those of the dynamical fields. Indeed, using the invariance of the kinetic Lagrangian and the Noether identity \eqref{eq:stueck noether} we find
\beq
\frac{\delta\Lag_m'}{\delta\sigma^\alpha}=
-\frac{\delta\Lag_m'}{\delta\varphi^i}\,R^i{}_{\beta}(S^{-1})^\beta_\alpha\,,
\eeq
which is only possible of course because $S^\beta_\alpha$ is (perturbatively) invertible.

\paragraph{Applications to the BRST-BV formalism.}

In the context of the BV deformation analysis, Eq.\ \eqref{eq:stueck sigma} 
implies that the cohomology of the differential $\gamma\,$
is trivial in strictly positive pureghost number $g\,$: 
$H^{g>0}(\gamma)\cong0\,$. Indeed, for the free theory we infer that
\beq
C^\alpha=\gamma(-\sigma^\alpha)\,,
\eeq
and hence the Stueckelberg ghosts (as well as all their derivatives) 
are always $\gamma$-exact. 
This leads to some interesting consequences that will become 
more explicit in the examples we analyze below. 
We will see, in particular, that any deformation of the gauge algebra of 
Stueckelberg transformations is necessarily trivial.

A simple but important theorem that follows from the property 
$H^{g>0}(\gamma)\cong0$ is that Stueckelberg gauge 
theories do not admit Chern--Simons-like vertices. 
Denote by $a^{0,D}_0$ the antifield-zero part of (the integrand of) 
the BV functional, at first order in deformation around the 
free theory. In $a^{0,D}_0$ will appear the possible cubic vertices 
to be added to the quadratic Lagrangian.
Then, if $a^{0,D}_0$ solves the master equation, we can always add to it a 
homogeneous solution $\bar{a}^{0,D}_0$ that satisfies
\beq \label{eq:w0bar}
\gamma\bar{a}{}^{0,D}_0 + d b^{1,D-1} = 0\,,
\eeq
where $b^{1,D-1}$ has ghost number 1 and form degree $D-1\,$. 
Such $\bar{a}{}^{0,D}_0$ terms correspond to vertices that are 
invariant, modulo a total derivative, under the gauge symmetries 
of the free theory. We can then identify two types of such vertices: 
Born--Infeld terms, which are exactly invariant, i.e.,  
$\gamma\bar{a}{}^{0,D}_0=0$; and Chern--Simons terms, 
for which $b^{1,D-1}\neq 0\,$. 
Of course, starting from a Born--Infeld term one can always 
integrate by parts, but the $b^{1,D-1}$ generated in this way 
will be $\gamma$-exact modulo $d\,$, 
due to the fact that $\gamma$ and $d$ anticommute. 
Thus, a true Chern--Simons vertex is 
defined by a $b^{1,D-1}$ that is nontrivial 
in the cohomology $H^{1,D-1}(\gamma|d)\,$ and such that 
$db^{1,D-1}$ is trivial in $H^{1}(\gamma)\,$.

Now, operate in equation \eqref{eq:w0bar} with $\gamma$ to get
$0=d(\gamma b^{1,D-1})$, which implies, by using the algebraic Poincar\'e
Lemma recalled in \cite{Barnich:2000zw}, that there exists 
$b^{2,D-2}$ such that 
$\gamma b^{1,D-1} + d b^{2,D-2} = 0\,$. 
Acting repeatedly with $\gamma\,$, one
produces the descent of equations
\begin{align}
\gamma \bar{a}{}^{0,D}_0 + d b^{1,D-1} &= 0\,, \\
\gamma b^{1,D-1} + d b^{2,D-2} &= 0\,, \\
&\vdots \nonumber \\
\gamma b^{k,D-k} &=0\;, \qquad k\leqslant D\;,
\end{align}
where the last equation is reached either because $k=D$ 
and a zero-form cannot be $d\,$-exact, or for $k<D\,$ in case 
one happens to reach an element $b^{k,D-k}$ in the 
cohomology of ${\gamma}\,$. 
But the last equation of the descent yields that 
$b^{k,D-k}=0$ in the cohomology because of the fact that 
$H^{g>0}(\gamma)\cong0$ in a Stueckelberg field theory, 
which then implies, working backwards along the 
ladder, that $b^{p,D-p}=0$ for all $p\,$. 
Therefore $\gamma\bar{a}^{0,D}_0=0\,$, 
which shows that any vertex that does not deform the 
gauge symmetry of a free Stueckelberg theory is necessarily 
of the Born--Infeld type. This result will prove very convenient, 
since in general Born--Infeld vertices are much easier to 
classify than Chern--Simons ones, and indeed we will be able to 
perform an exhaustive analysis for the example theories we study.


\section{Massive Yang--Mills theory} \label{sec:massiveym}

Our first example is the massive version of Yang--Mills theory in the Stueckelberg formulation. The starting point is the free 
theory of a collection\footnote{The color indices $a, b, \ldots$ run over 
a finite number $n_v$ of values. In our conventions, they are raised 
and lowered using an Euclidean metric $\delta_{ab}$ in the internal space of vector fields.} of massive spin-1 fields $A^a_{\mu}$,
\beq
S_0=\int d^Dx\, \left[-\tfrac{1}{4}\,F_a^{\mu\nu}F^a_{\mu\nu}-\tfrac{1}{2}\,\D_{\mu}\pi_a\D^{\mu}\pi^a\right]\,,
\eeq
where the Stueckelberg scalars are $\pi^a$, and $F^a_{\mu\nu}\coloneqq\partial_{\mu}A^a_{\nu}-\partial_{\nu}A^a_{\mu}$ will 
always denote the field strength of the free theory. We have also defined
\beq
\D_{\mu}\pi^a\coloneqq\partial_{\mu}\pi^a-mA^a_{\mu}\,,
\eeq
and note that all the fields are assumed to have the same mass $m$. The above action enjoys the gauge symmetry
\beq
\delta_{\epsilon}A^a_{\mu}=\partial_{\mu}\epsilon^a\,,\qquad \delta_{\epsilon}\pi^a=m\epsilon^a\,.
\eeq
The choice of unitary gauge $\pi^a=0$ then returns the usual free Proca action for the spin-1 fields $A^a_\mu$; see section 
\ref{subsec:stueck}.

To construct the BV functional we introduce ghosts $C^a$, equal in number to the gauge symmetries, and antifields $A^{*\mu}
_a$ and $C^{*}_a$. From \eqref{eq:bv bvfunc} we then have, at quadratic order,
\beq
\label{3.4}
W_0=S_0+\int d^Dx \Big[A^{*\mu}_a\partial_{\mu}C^a+m\pi^{*}_aC^a\Big]\,,
\eeq
which can be checked to satisfy the master equation $(W_0,W_0)=0\,$.

The BRST differential (of the free theory) is decomposed as $s=\gamma+\delta$, where
\beq
\gamma A^a_{\mu}=\partial_{\mu}C^a\,,\qquad \gamma \pi^a=mC^a\,,
\eeq
\beq
\delta A^{*\mu}_a=\frac{\delta S_0}{\delta A^a_{\mu}}=\partial_{\nu}F_a^{\nu\mu}-m^2A_a^{\mu}+m\partial^{\mu}\pi_a\,,\qquad 
\delta\pi^{*}_a=\frac{\delta S_0}{\delta \pi^a}=\Box\pi_a-m\partial_{\mu}A_a^{\mu}\,,
\eeq
\beq
\delta C^{*}_a=-\partial_{\mu}A^{*\mu}_a+m\pi^{*}_a\,,
\eeq
and the action of $\gamma$ and $\delta$ on the variables not shown is by definition zero.

\subsection{Cubic deformations}
\label{subsec3.1}

We write $W_1$ as
\beq
W_1=\int d^Dx\, (a_0+a_1+a_2)\,,
\label{first-order-def}
\eeq
where ${\rm antifld}(a_k)=k$; the reason why the expansion stops 
at antifield number 2 is explained in \cite{Boulanger:2000rq}
and results from the general theorems of \cite{Barnich:1994db}
that can be applied to the local, irreducible  theory \eqref{3.4}.
Referring to Eq.\ \eqref{eq:bv bvfunc}, the interpretation of the 
various terms $a_i$ in antifield numbers 0, 1 and 2 is as follows. 
The local scalar density $a_0$ encodes the infinitesimal deformations of 
the Lagrangian, 
while the local scalar density $a_1$ encodes the first-order deformation 
of the gauge transformation laws. Finally, $a_2$ gives the information 
about the first-order deformations of the Abelian gauge algebra. 
The equation $sW_1=0$ can now be decomposed with respect to the antifield number, with the 
results \cite{Barnich:1994db}
\beq
\left\{
\bal \label{eq:ym eqs_a}
\gamma a_0+\delta a_1+\td&=0\,,\\
\gamma a_1+\delta a_2+\td&=0\,,\\
\gamma a_2 &=0\,.
\eal
\right.
\eeq

We assume that the deformation starts at cubic order
and modifies the gauge algebra of the free theory. Therefore, we write
\beq \label{eq:ym a2}
a_2= \tfrac{g}{2}\,f^a{}_{bc}C^{*}_aC^bC^c\,,
\eeq
where $f^a{}_{bc}=f^a{}_{[bc]}$ is an otherwise arbitrary (at this stage) 
constant tensor. This is the most general expression at 
lowest order in derivatives. Obviously $\gamma a_2=0$, in agreement with 
the last equation in \eqref{eq:ym eqs_a}, but in fact 
$a_2$ is moreover $\gamma$-exact, 
\beq
a_2=\gamma\left(\tfrac{g}{2m}\,f^a{}_{bc}C^{*}_a\pi^bC^c\right)\,,
\eeq
which follows because $C^a=\gamma(\pi^a/m)$. This is the first instance of the phenomenon we anticipated in section 
\ref{subsec:stueck}: the fact that the ghosts $C^a$ do not belong to the cohomology $H(\gamma)$ implies a trivial $a_2$, and 
therefore we conclude that there is no nontrivial deformation of the gauge algebra. Note that this remains true if one considers 
higher derivative contractions, and will also hold at all orders in perturbations. 
This is in contrast to what occurs in the massless 
case \cite{Barnich:1993pa,Barnich:1994mt}, 
where the term in \eqref{eq:ym a2} is precisely the starting point that 
leads to Yang--Mills theory.

It should be emphasized that the triviality of the above $a_2$ does not 
mean we cannot use it. The $\gamma$-exactness of 
$a_2$ implies that there exists a choice of field-dependent gauge parameters
for which $a_2=0$, but such choice is not necessarily the smartest 
one. Indeed, here we are interested in seeing explicitly how the deformation 
analysis is modified in the presence of mass 
terms, and hence it is useful to keep the same point of departure as in the 
massless case. 
We will have more to say about this aspect below, 
in section \ref{subsec:triviality}, where we make explicit 
the field redefinition that 
maps a solution constructed starting from \eqref{eq:ym a2} to the one where 
both $a_2$ and $a_1$ are taken to be zero.
The approach we follow in this section is dictated by the wish to reproduce
the Yang--Mills theory in the unitary gauge followed by the massless limit.
Notice that a similar guiding principle was followed in 
\cite{Zinoviev:2011fv,Zinoviev:2014zka,Zinoviev:2016tua}.

With this discussion in mind, we continue the calculation to 
get $a_1$ from the second equation in \eqref{eq:ym eqs_a}. 
First we need
\beq\bal \label{eq:ym delta_a2}
\delta a_2&= gf^a{}_{bc}A^{*\mu}_a\partial_{\mu}C^bC^c + \tfrac{g}{2}\,f^a{}_{bc}\pi^{*}_a(mC^b)C^c+\td\\
&= - \gamma\left(gf^a{}_{bc}A^{*\mu}_aA^b_{\mu}C^c+\tfrac{g}{2}\,f^a{}_{bc}\pi^{*}_a\pi^bC^c\right)+\td\,,
\eal\eeq
and to obtain the second line we note that, given any two Grassmann-odd 
variables $\alpha$ and $\beta$, one has 
$\gamma(\alpha\beta)=(\gamma\alpha)\beta-\alpha(\gamma\beta)$. 
Here we encounter another subtle point that is 
characteristic of the Stueckelberg formulation: 
the expression $\partial_{\mu}C^a$ can be 
written both as $\gamma A^a_{\mu}$ 
or as $\gamma(\partial_{\mu}\pi^a/m)$, meaning that our $a_2$ will admit two 
independent ``liftings''. We have checked 
however that the second option leads to a cubic vertex identically 
null, $a_0 \equiv 0\,$, hence we discard it. 
Using the result of \eqref{eq:ym delta_a2} in the second line of 
\eqref{eq:ym eqs_a}, we obtain the particular solution
\beq
a_1=gf^a{}_{bc}A^{*\mu}_aA^b_{\mu}C^c+\tfrac{g}{2}\,f^a{}_{bc}\pi^{*}_a\pi^bC^c\,.
\eeq
From this we can read off the order-$g$ deformation of the gauge symmetry:
\beq
\delta_{\epsilon}^{(1)}A^a_{\mu}=gf^a{}_{bc}A^b_{\mu}\epsilon^c\,,\qquad \delta_{\epsilon}^{(1)}\pi^a=\tfrac{g}{2}\,f^a{}_{bc}
\pi^b\epsilon^c\,.
\eeq
Note that to the above $a_1$ we may add a solution to the homogeneous 
equation $\gamma a_1+\td=0$, which we denote by 
$\bar{a}_1\,$. In order to stay as close as possible to the 
massless spin-1 case 
for which there exists no such $\bar{a}_1\,$'s, we choose not 
to take any such representatives, which anyway are trivial in the 
cohomology (recall that $H^g(\gamma)\cong0$ for $g>0$)
and hence can be eliminated by a field redefinition.\footnote{For example, 
$\bar{a}_1=A^{*\mu}_a {\cal D}_\mu\pi^b C^c f^a{}_{bc}$ is a candidate
that is readily seen to be $\gamma\,$-exact.}

Finally, to find $a_0$ we compute
\beq\bal
\delta a_1&=-gf^a{}_{bc} \left[ \tfrac{1}{2}\,F_a^{\mu\nu}F^b_{\mu\nu}C^c+m^2A_a^{\mu}A^b_{\mu}C^c+F_a^{\mu\nu}A^b_{\nu}
\partial_{\mu}C^c-\partial^{\mu}\pi_aA^b_{\mu}(mC^c)\right.\\
&\quad+\left.\tfrac{1}{2}\,\partial^{\mu}\pi_a\partial_{\mu}\pi^bC^c+\tfrac{1}{2}\,\partial^{\mu}\pi_a\pi^b\partial_{\mu}C^c+\tfrac{1}{2}\,
\partial_{\mu}A_a^{\mu}\pi^b(mC^c)\right]+\td\,.
\eal\eeq
This is $\gamma$-exact without a priori
imposing $f_{abc}\coloneqq \delta_{ad} f^d{}_{bc}=f_{[abc]}$. Indeed we obtain
\beq\bal
\delta a_1 + \td &=-\gamma\left(-\tfrac{g}{2}\,f^a{}_{bc}F_a^{\mu\nu}A^b_{\mu}A^c_{\nu}-\tfrac{g}{2}\,f^a{}_{bc}A_a^{\mu}\partial_{\mu}
\pi^b\pi^c - g\,f^a{}_{bc} \partial^\mu \pi_a A_\mu^b \pi^c \right.\\
&\quad\left. + \tfrac{g}{2m} f^a{}_{bc} F^{\mu\nu}_a F_{\mu\nu}^b \pi^c + gm f^a{}_{bc} A^\mu_a A_\mu^b \pi^c + \tfrac{g}{2m} f^a{}_{bc} \partial_\mu \pi_a \partial^\mu \pi^b \pi^c \right)\,,
\eal\eeq
implying the consistency at first order in $g$ of the ``exotic'' vertex
\beq\bal
a^{\rm exotic}_0 &= -\tfrac{g}{2}\,f^a{}_{bc}F_a^{\mu\nu}A^b_{\mu}A^c_{\nu}-\tfrac{g}{2}\,f^a{}_{bc}A_a^{\mu}\partial_{\mu}
\pi^b\pi^c - g\,f^a{}_{bc} \partial^\mu \pi_a A_\mu^b \pi^c \\
&\quad+ \tfrac{g}{2m} f^a{}_{bc} F^{\mu\nu}_a F_{\mu\nu}^b \pi^c + gm f^a{}_{bc} A^\mu_a A_\mu^b \pi^c + \tfrac{g}{2m} f^a{}_{bc} \partial_\mu \pi_a \partial^\mu \pi^b \pi^c\,.
\eal\eeq

However, since our guiding principle is that, in the unitary gauge 
followed by massless limit, one should recover massless Yang--Mills theory, 
one must finally impose the relation $f_{abc}=f_{[abc]}\,$. 
As a matter of fact, the first term of $a^{\rm exotic}_0$ is the right cubic 
part of the Yang--Mills Lagrangian, only when the structure constants 
satisfy $f_{abc}=f_{[abc]}\,$. We thus 
arrive at the result
\beq \label{eq:ym a0result}
a_0=\tfrac{g}{2}\,f^a{}_{bc}\left(-F_a^{\mu\nu}A^b_{\mu}A^c_{\nu}
+A_a^{\mu}\partial_{\mu}\pi^b\pi^c\right)\,,
\quad f_{abc} = f_{[abc]}\;,
\eeq
which corresponds to the cubic vertices of the deformed theory. 
We can again envisage adding homogeneous solutions 
satisfying $\gamma\bar{a}_0+\td=0\,$. 
From the theorem given in section \ref{subsec:stueck}, 
we know that all such solutions must be of the Born--Infeld type, 
i.e. solution of $\gamma\bar{a}_0=0\,$. 
The cohomology of $\gamma$ in the space of fields and their derivatives 
is generated by\footnote{Note that the variables $[F^a_{\mu\nu}]$ and 
$[\D_{\mu}\pi^a]$ are not linearly independent due to the identity 
$2\partial_{[\mu}\D_{\nu]}\pi^a + m F^a_{\mu\nu} \equiv 0\,$. A basis 
can be obtained by taking the set $\{\partial_{(\mu_1}\ldots \partial_{\mu_{n-1}}F^a{}_{\mu_n)\nu} \;, \partial_{(\mu_1}\ldots 
\partial_{\mu_{n-1}}\D_{\mu_n)}\pi^a \}\,$, $n=1,2,\ldots\,$.}
\beq
H^0(\gamma)\cong\left\{f([F^a_{\mu\nu}],[\D_{\mu}\pi^a])\right\}\,.
\eeq
It is easy to see that there exists no two-derivative cubic 
invariant contractions built out of these elements. The simplest ones 
are the three-derivative vertices
\beq\bal \label{eq:ym a0bar}
\bar{a}_0^{(1)}&=f^a{}_{bc}F_a^{\mu\nu}{\mathcal{D}}_{\mu}\pi^b{\mathcal{D}}_{\nu}\pi^c\,,\\
\bar{a}_0^{(2)}&=f^a{}_{bc}F_a^{\mu\nu}F^b_{\phantom{b}\nu\rho}F^{c\rho}_{\phantom{c\rho}\mu}\,.\\
\eal\eeq
Note that such Born--Infeld terms, when taken alone, automatically 
solve the master equation and hence cannot generate an 
obstruction at higher orders in perturbations --- although they will 
get Yang--Mills covariantized, meaning that at the end of the deformation 
procedure they become Born--Infeld terms under the non-linear completion 
of the gauge symmetries. For this reason it is consistent to ignore them for now, although we will consider such deformations later.

\subsection{Quartic deformations}

It is easy to see that the antibracket of $W_1$ with itself takes the form
\beq
(W_1,W_1)=\int d^Dx\, (\alpha_0+\alpha_1+\alpha_2)\,,
\eeq
where again ${\rm antifld}(\alpha_k)=k$. We find
\beq\bal \label{eq:ym alphas1}
\alpha_2&=-g^2f^a{}_{b[c}f_{|a|de]}C^{*b}C^cC^dC^e\,,\\
\alpha_1&=3g^2f^a{}_{b[c}f_{|a|de]}A^{*b\mu}A^c_{\mu}C^dC^e+g^2f^a{}_{b[c}f_{|a|d]e}\pi^{*b}\pi^cC^dC^e\,,\\
\alpha_0&=-\tfrac{3g^2}{2}\,f^a{}_{b[c}f_{|a|de]}F^b_{\mu\nu}A^{c\mu}A^{d\nu}C^e+g^2f^a{}_{bc}f_{ade}A^{b\mu}A^{c\nu}A^d_{\mu}
\partial_{\nu}C^e\\
&\quad+\tfrac{g^2}{2}\,f^a{}_{bc}f_{ade}\Big(2\partial^{\mu}\pi^b\pi^cA^d_{\mu}+\partial_{\mu}A^{b\mu}\pi^c\pi^d+2A^{b\mu}
\partial_{\mu}\pi^c\pi^d\Big)C^e+\td\,.
\eal\eeq
As before we decompose the BV functional $W_2$ with respect to the antifield number,
\beq
W_2=\int d^Dx\, (b_0+b_1+b_2)\,,
\eeq
so that the order $g^2$ master equation $sW_2+\tfrac{1}{2}(W_1,W_1)=0$ yields
\beq
\left\{
\bal \label{eq:ym eqs_b}
\gamma b_0+\delta b_1+\tfrac{1}{2}\,\alpha_0+\td&=0\,,\\
\gamma b_1+\delta b_2+\tfrac{1}{2}\,\alpha_1+\td&=0\,,\\
\gamma b_2+\tfrac{1}{2}\,\alpha_2+\td&=0\,.\\
\eal
\right.
\eeq
A priori the structure constants do not have to satisfy the Jacobi identity since the last equation can be solved for $b_2$ for general structure constants $f_{abc}$ (which would not be possible in the massless case because 
$C^{*b}C^cC^dC^e$, being now in $H(\gamma)\,$, would yield an obstruction). 
Explicitly we get
\begin{equation}
\label{solb2}
    b_2 = \tfrac{g^2}{2m} f^a{}_{b[c} f^b{}_{de]} C^*_a \pi^c C^d C^e\,,
\end{equation}
up to a solution $\bar{b}_2$ of the homogeneous equation 
$\gamma b_2 = 0\,$. 
Again, because $H^g(\gamma)$ is trivial in strictly positive pureghost number, 
we have the choice to discard such possibilities.
Moreover, the $\bar{b}_2$'s that exist for the massive theory would bring 
too many derivatives in the Lagrangian.

Moving on to the next term, from \eqref{eq:ym alphas1} and \eqref{solb2} one can calculate
\begin{equation}
    - \delta b_2 - \tfrac{1}{2} \alpha_1 + \td = \dots - \tfrac{g^2}{8} f^a{}_{b[c} f^b{}_{de]} \pi^*_a \pi^c C^d C^e\,.
\end{equation}
The presence of the last term makes it impossible to solve the second equation 
in \eqref{eq:ym eqs_b} for $b_1\,$, since it cannot be produced by 
$\gamma b_1$ due to the antisymmetrization of color indices. 
It is therefore at this stage that we finally have to impose the Jacobi identity,
\begin{equation}
\label{Jacobi1}
    f^a{}_{b[c} f^b{}_{de]} = 0\;, 
\end{equation}
so that $b_2$ vanishes, as in the Yang--Mills case. 
A useful equivalent form of \eqref{Jacobi1} is
\begin{equation}
\label{Jacobi2}
    f^a{}_{b[c}f_{|a|d]e}=-\tfrac{1}{2}\,f^a{}_{cd}f_{abe}\,.
\end{equation}

Coming back to the part with antifield number one in \eqref{eq:ym alphas1} and using \eqref{Jacobi1} and \eqref{Jacobi2} we can calculate
\beq\bal
\alpha_1&=-\tfrac{g^2}{2}\,f^a{}_{be}f_{acd}\pi^{*b}\pi^cC^dC^e\\
&=\gamma\left(-\tfrac{g^2}{6m}\,f^a{}_{bc}f_{ade}\pi^{*b}\pi^c\pi^dC^e\right)\,.\\
\eal\eeq
Since now $b_2=0$ by virtue of the Jacobi identity, the second equation in \eqref{eq:ym eqs_b} implies
\beq
b_1=\tfrac{g^2}{12m}\,f^a{}_{bc}f_{ade}\pi^{*b}\pi^c\pi^dC^e\,,
\eeq
which results in the following second-order deformation of the gauge symmetry:
\beq
\delta_{\epsilon}^{(2)}A^a_{\mu}=0\,,\qquad \delta_{\epsilon}^{(2)}\pi^a=\tfrac{g^2}{12m}\,f^a{}_{bc}f^c{}_{de}\pi^b\pi^d\epsilon^e\,.
\eeq

The final step is to find $b_0\,$ from the first equation in \eqref{eq:ym eqs_b}. After some manipulations $\alpha_0$ simplifies to
\beq\bal
\alpha_0&=2g^2f^a{}_{bc}f_{ade}A^{b\mu}A^{c\nu}A^d_{\mu}\partial_{\nu}C^e
-\tfrac{g^2}{2}\,f^a{}_{bc}f_{ade}A^{b\mu}
\pi^c\pi^d\partial_{\mu}C^e\\
&\quad+\tfrac{g^2}{2}\,f^a{}_{be}f_{acd}A^{b\mu}\partial_{\mu}\pi^c\pi^dC^e+\td\,.
\eal\eeq
We observe that the first term gives the quartic Yang--Mills vertex,
\beq
2g^2f^a{}_{bc}f_{ade}A^{b\mu}A^{c\nu}
A^d_{\mu}(\gamma A^e_{\nu})=\gamma\left(\tfrac{g^2}{2}\,
f^a{}_{bc}f_{ade}A^{b\mu}A^{c\nu}
A^d_{\mu}A^e_{\nu}\right)\,.
\eeq
Expanding $\delta b_1$ and collecting terms we get
\beq\bal
\tfrac{1}{2}\,\alpha_0+\delta b_1&=\gamma\left(\tfrac{g^2}{4}\,
f^a{}_{bc}f_{ade}A^{b\mu}A^{c\nu}A^d_{\mu}A^e_{\nu}\right)+
\tfrac{g^2}{3}\,f^a{}_{b(d}f_{|ac|e)}A^{b\mu}\partial_{\mu}\pi^c\pi^dC^e\\
&\quad-\tfrac{g^2}{6}\,f^a{}_{bc}f_{ade}A^{b\mu}\pi^c\pi^d\partial_{\mu}
C^e-\tfrac{g^2}{12m}\,f^a{}_{bc}f_{ade}\partial^{\mu}
\pi^b\pi^c\partial_{\mu}\pi^dC^e\\
&\quad-\tfrac{g^2}{12m}\,f^a{}_{bc}f_{ade}\partial^{\mu}\pi^b\pi^c\pi^d\partial_{\mu}C^e+\td\;.
\eal\eeq
By using a general Ansatz for $b_0\,$,
\beq\bal
b_0&=-\tfrac{g^2}{4}\,f^a{}_{bc}f_{ade}A^{b\mu}A^{c\nu}A^d_{\mu}A^e_{\nu}+x_1g^2f^a{}_{bc}f_{ade}A^{b\mu}\pi^c\pi^dA^e_{\mu}+
\tfrac{x_2g^2}{m}\,f^a{}_{bc}f_{ade}A^{b\mu}\pi^c\partial_{\mu}\pi^d\pi^e\\
&\quad+\tfrac{x_3g^2}{m^2}\,f^a{}_{bc}f_{ade}\partial^{\mu}\pi^b\pi^c\partial_{\mu}\pi^d\pi^e\;,
\eal\eeq
and substituting in \eqref{eq:ym eqs_b}, we find that 
$x_1=0$, $x_2=-1/6$ and $x_3=1/24$, so that the final result for the 
quartic vertices of the theory is given by
\beq \label{eq:ym b0result}
b_0=-\tfrac{g^2}{4}\,f^a{}_{bc}f_{ade}A^{b\mu}A^{c\nu}A^d_{\mu}A^e_{\nu}-\tfrac{g^2}{6m}\,f^a{}_{bc}f_{ade}A^{b\mu}
\pi^c\partial_{\mu}\pi^d\pi^e+\tfrac{g^2}{24m^2}\,f^a{}_{bc}f_{ade}\partial^{\mu}\pi^b\pi^c\partial_{\mu}\pi^d\pi^e\,.
\eeq

We could also consider adding the quartic solutions of the homogeneous equation $\gamma \bar{b}_0 = 0\,$, which are of the schematic form $F^4$, $F^2(\D\pi)^2$ and $(\D\pi)^4$, with the color indices being contracted with either a pair of structure constants or with $\delta_{ab}$. We omit the list of all such contractions as we will have no occasion to use them, but the task is of course straightforward.

\subsection{Comparison with the full nonlinear theory}

It is instructive to check our results by expanding the full non-linear model 
obtained by performing a Stueckelberg replacement 
of massive Yang--Mills theory, see e.g.\ \cite{Hinterbichler:2011tt,Goon:2014ika}.
To this end it is helpful to switch to matrix notation and write 
${\bf A}_{\mu}=A^a_{\mu}T_a$, with $T_a$ the generators of the gauge group 
normalized such that
\beq
[T_a,T_b]=if^c{}_{ab}T_c\,.
\eeq
The action is
\beq
S=\int {d^Dx} \left[-\tfrac{1}{2}\,{\rm tr}({\bf F}^{\mu\nu}{\bf F}_{\mu\nu})-\tfrac{m^2}{2{\cal T}}\,{\rm tr}({\bf A}^{\mu}{\bf A}_{\mu})\right]\,,
\eeq
where ${\bf F}_{\mu\nu}$ is the Yang--Mills field strength and ${\cal T}$ is 
the Dynkin index of the fundamental representation 
of the group, defined via ${\rm tr}(T_aT_b)={\cal T}\delta_{ab}$. 
We introduce the Stueckelberg fields $\pi^a$ mimicking the 
gauge transformation under which the kinetic term remains invariant,
\beq
{\bf A}_{\mu}\to{\bf A}_{\mu}'=U{\bf A}_{\mu}U^{-1}-\tfrac{i}{g}\,U\partial_{\mu}U^{-1}\,,\qquad U\coloneqq 
e^{ig\pi^aT_a/m}\,.
\eeq
Let us ignore the pure vector part of the action and focus on 
the new operators involving the Stueckelberg modes,
\beq
\Lag'=-\tfrac{m^2}{2{\cal T}}\left[\tfrac{2i}{g}\,{\rm tr}({\bf A}^{\mu}U^{-1}\partial_{\mu}U)+\tfrac{1}{g^2}\,{\rm tr}(\partial^{\mu}
U\partial_{\mu}U^{-1})\right]\,.
\eeq
Both terms in $\Lag'$ are functions of the matrix
\beq \label{eq:ym udu}
U^{-1}\partial_{\mu}U=\tfrac{ig}{m}\,\partial_{\mu}\pi^a\ell_{a}{}^b(\pi)T_b\,,
\eeq
where we defined
\beq \label{eq:ym vielbein}
\ell_{a}{}^b(\pi)\coloneqq\delta_{a}^{b}+\sum_{k=1}^{\infty}\tfrac{(g/m)^k}{(k+1)!}(\pi^{b_1}\cdots\pi^{b_k})(f_{b_1a}^{\phantom{b_1a}c_1}
f_{b_2c_1}^{\phantom{b_2c_1}c_2}\cdots f_{b_kc_{k-1}}^{\phantom{b_kc_{k-1}}b})\,.
\eeq
It is straightforward to check that $\ell_{a}{}^{b}$ satisfies the differential equation
\beq
\frac{\partial\ell_{a}{}^{c}}{\partial\pi^b}-\frac{\partial\ell_{b}{}^{c}}{\partial\pi^a}=-\tfrac{g}{m}\,f_{de}{}^c\ell_{a}{}^{d}\ell_{b}{}^{e}\,,
\eeq
which can be used as an alternative definition as was done in \cite{Zinoviev:2006im}. Expressing $\ell_{a}{}^{b}$ in closed form is not possible for generic gauge groups, but for the particular case of $SU(2)$ we can perform the sum to find
\beq
\ell_{ab}=\delta_{ab}\,\frac{\sin\left(\tfrac{g}{m}\sqrt{\pi^2}\right)}{\tfrac{g}{m}\sqrt{\pi^2}}+\frac{\pi_a\pi_b}{\pi^2}\left(1-
\frac{\sin\left(\tfrac{g}{m}\sqrt{\pi^2}\right)}{\tfrac{g}{m}\sqrt{\pi^2}}\right)-\frac{\epsilon_{abc}\pi^c}{\sqrt{\pi^2}}\left(\frac{1-
\cos\left(\tfrac{g}{m}\sqrt{\pi^2}\right)}{\tfrac{g}{m}\sqrt{\pi^2}}\right)\,,
\eeq
with $\pi^2\coloneqq \delta_{ab}\pi^a\pi^b$.

Returning to the general case, we substitute \eqref{eq:ym udu} into the Lagrangian to obtain the Stueckelberg part of the 
action,
\beq
\label{fullsolution}
\Lag'=m\partial^{\mu}\pi^a\ell_{a}{}^{b}(\pi)A_\mu^{b}-\tfrac{1}{2}\,\gamma_{ab}(\pi)\partial^{\mu}\pi^a\partial_{\mu}\pi^b\,,
\eeq
where $\gamma_{ab}\coloneqq\ell_{a}{}^{c}\ell_{bc}$, which gives the interpretation 
of $\ell_{a}{}^{b}$ as a field space vielbein \cite{Slavnov:1971mz}.
It is now direct to expand this expression using the explicit result 
\eqref{eq:ym vielbein} to recover the cubic and quartic vertices we 
obtained through the deformation procedure.

\subsection{Decoupling limit}

The quartic vertices we have derived, Eq.\ \eqref{eq:ym b0result}, are clearly singular in the massless limit, i.e.\ we cannot 
simply take $m\to0$. A more sensible thing to do with an interacting theory is to study the decoupling limit, that is to consider 
the limiting value of the coupling constants in such a way that (1) the number of degrees of freedom is preserved, (2) the 
Stueckelberg fields become gauge invariant and hence physical, 
carrying propagating degrees of freedom rather than pure 
gauge ones, 
and (3) the resulting theory retains some nonlinearities and is therefore nontrivial. In practice we can achieve 
these requirements by identifying the smallest energy scale---let's call it $M$---that plays a role in the theory, and to take the 
limit of the parameters of the model such that $M$ is kept fixed but all other scales that are greater than $M$ go to infinity.

By inspection of the vertices and modified gauge transformations, it is clear that the decoupling limit of massive YM theory is 
achieved by letting $g,m\to0$ with the scale $M\coloneqq(m/g)^{2/(D-2)}$ kept finite.\footnote{Note that in $D$ dimensions the 
coupling constant $g$ has mass dimension $-(D-4)/2$ (since the canonically normalized fields have dimension $(D-2)/2$). The 
scale $M$ thus carries mass dimension one.} The resulting action is
\beq
S=\int d^Dx \left[-\tfrac{1}{4}\,F^{a}_{\mu\nu}F_a^{\mu\nu}-\tfrac{1}{2}\,\partial_{\mu}\pi_a\partial^{\mu}\pi^a+\tfrac{1}{24M^{D-2}}\,f^{a}{}_{bc}
f_{ade}\partial_{\mu}\pi^b\pi^c\partial^{\mu}\pi^d\pi^e+\cdots\right]\,,
\eeq
and the omitted terms correspond to higher order scalar vertices suppressed by increasing powers of $M$. The appearance of 
a dimensionful parameter (even though the fields are massless) means that the theory of the scalar sector is non-renormalizable and must be regarded as an EFT with UV cutoff $M$ \cite{Hinterbichler:2011tt}. In the decoupling limit the scalar and vector sectors are truly decoupled from one another (unlike what happens for instance in massive gravity); the vector part of the action is nothing but a sum of Maxwell terms, meaning that we must necessarily lose the Yang--Mills interactions when taking this limit, but the scalar self-interaction on the other hand are nontrivial and correspond to a non-linear sigma model. Indeed, in this decoupling limit the gauge symmetry reduces to
\beq
\delta_{\epsilon}A^a_{\mu}=\partial_{\mu}\epsilon^a\,,\qquad \delta_{\epsilon}\pi^a=0\,,
\eeq
consistent with the expectation that in this limit the vector fields become massless, with the longitudinal modes now being 
physically propagated by the scalars $\pi^a$.

\subsection{Triviality of the gauge algebra}
\label{subsec:triviality}

\paragraph{Determination of the field redefinition.}
We have seen that, in Stueckelberg gauge theories, any deformation of the 
gauge transformation laws and of the gauge algebra --- encoded respectively 
in the terms $a_1$ and $a_2$ for the first order deformation of the BV 
functional --- are ultimately trivial by virtue of the fact that the ghosts 
are $\gamma$-exact. In this subsection we make explicit the field redefinition 
and field-dependent gauge parameter redefinition that provide 
the mapping from the solution of the first-order deformation 
\eqref{first-order-def}, which we solved for in section \ref{subsec3.1}, 
to the trivial solution, modulo Born--Infeld type vertices, defined by
\begin{align}
    a_2^{t} &= \gamma (\tfrac{g}{2m} f^a_{\phantom{a}bc} C^*_a \pi^b C^c)\;,\\
    a_1^{t} &= \delta (\tfrac{g}{2m} f^a_{\phantom{a}bc} C^*_a \pi^b C^c)\;, \\
    a_0^{t} &= 0\;.
\end{align}
Since
\begin{equation}
\label{trivgaugalgspin1}
a_2 = \tfrac{g}{2} f^a_{\phantom{a}bc} C^*_a C^b C^c 
= \gamma (\tfrac{g}{2m} f^a_{\phantom{a}bc} C^*_a \pi^b C^c) = a_2^{t}\;,
\end{equation}
the resolution of $\delta a_2 + \gamma a_1 = \td$ has to give the same solution for $a_1$ and $a_1^{t}$ modulo $
\gamma$-closed terms and modulo total derivatives, 
i.e.\ $\gamma \left( a_1^{t} - a_1 \right) = \partial_\mu j^\mu\,$. 
Because of the triviality of the cohomology of $\gamma$ modulo $d$ in 
strictly positive pureghost number, 
$a_1^{t} - a_1$ has to be $\gamma$-exact modulo a total derivative. 
Thus the goal is to solve for $c_1$ in 
\begin{equation}
\label{diffa1gammaX}
\delta (\tfrac{g}{2m} f^a_{\phantom{a}bc} C^*_a \pi^b C^c) - g f^a_{\phantom{a}bc} A^{*\mu}_a A_\mu^b C^c - \tfrac{g}{2} 
f^a_{\phantom{a}bc} \pi^*_a \pi^b C^c = \gamma c_1 + \td\,.
\end{equation}
The knowledge of $c_1$ will give us the field redefinition connecting the two solutions.
Noting that the left-hand side of \eqref{diffa1gammaX} can be written as
\begin{equation}
\td + \tfrac{g}{2m} f^a_{\phantom{a}bc} A^{*\mu}_{\phantom{*}a} \left( \partial_\mu \pi^b C^c + \pi^b \partial_\mu C^c \right) - g 
f^a_{\phantom{a}bc} A^{*\mu}_{\phantom{*}a} A_\mu^b C^c\;,
\end{equation}
one finds that the solution for $c_1$ is provided by
\begin{equation}
c_1 = \tfrac{g}{2m^2} f^a_{\phantom{a}bc} A^{*\mu}_{\phantom{*}a} \pi^b \left( \partial_\mu \pi^c - 2m A_\mu^c \right)\,.
\end{equation}

In conclusion, the field redefinition at cubic order that enables us 
to go from the cubic vertex $a_0$ to the trivial cubic vertex 
$a_0^t=0\,$, up to strictly gauge-invariant vertices, 
is given by
\begin{equation}
\label{ExplFR}
A_\mu^a \longrightarrow A_\mu^a + \tfrac{g}{2m^2} f^a_{\phantom{a}bc} \pi^b \left( \partial_\mu \pi^c - 2m A_\mu^c \right)\,,
\end{equation}
and the redefinition of the gauge parameters that trivializes the cubic deformation of the gauge algebra can be read in \eqref{trivgaugalgspin1} as
\beq
\epsilon^a \longrightarrow \epsilon^a + \tfrac{g}{2m} f^a_{\phantom{a}bc} \pi^b \epsilon^c \,.
\eeq

\paragraph{Determination of the non-trivial cubic deformation.}

From the resolution of $\delta a_1 + \gamma a_0 = \td$ and \eqref{diffa1gammaX} 
we have that the difference between the two solutions $a_0$ and 
$a_0^{t}=0$ summed with $\delta c_1$ is $\gamma$-closed modulo a total derivative. 
Indeed,
\begin{align}\label{constrBI}
\gamma \left( a_0 - a_0^t + \delta c_1 \right) &= \gamma a_0 + \gamma \delta c_1 = - \delta \left( a_1 + \gamma 
c_1 \right) + \td
\nonumber \\
& = -\delta \left( \delta ( \tfrac{g}{2m} f^a_{\phantom{a}bc} C^*_a \pi^b C^c ) \right) + \td = \td\; .
\end{align}
Since a pureghost-zero solution of $\gamma a_0 + \td = 0$ is equivalent to 
$\gamma \bar{a}_0 = 0\,$, as we have seen in section \ref{subsec:stueck}, 
one can add total derivatives to the $\gamma$ modulo $d$ 
cocycle $a_0 - a_0^t + \delta c_1$ so that 
it becomes strictly annihilated by $\gamma\,$. 
Therefore we have
\begin{equation}\label{bara0delta}
\bar{a}_0 = a_0 + \delta c_1\;.
\end{equation}
Explicitly, the cocycle $\bar{a}_0$ reads
\begin{equation}
\bar{a}_0 = -\tfrac{g}{2} f^a_{\phantom{a}bc} F_a^{\mu\nu} A_\mu^b A_\nu^c + \tfrac{g}{m} f^a_{\phantom{a}bc} F_a^{\mu\nu} 
\partial_\mu \pi^b A_\nu^c - \tfrac{g}{2m^2} f^a_{\phantom{a}bc} F_a^{\mu\nu} \partial_\mu \pi^b \partial_\nu \pi^c\;.
\end{equation}
To make manifest that it belongs to $H^0(\gamma)\,$, we rewrite it as follows:
\begin{equation}
\bar{a}_0 = -\tfrac{g}{2m^2} f^a_{\phantom{a}bc} F_a^{\mu\nu} {\cal D}_\mu 
\pi^b {\cal D}_\nu \pi^c \;.
\end{equation}
This is actually the most general deformation which mixes 
$A_\mu^a$ and $\pi^a\,$; see \eqref{eq:ym a0bar}. 
Indeed, this $\bar{a}_0$ is the most general cubic term in 
$H^0(\gamma)\,$, at ghost number zero, that involves both 
$A_\mu^a$ and $\pi^a\,$.

In conclusion, the cubic deformation found in section \ref{subsec3.1} 
is obtained from the deformation of action
\begin{equation}
\nonumber
S_0 = \int d^Dx \left( -\tfrac{1}{4} F_{\mu\nu}^a F_a^{\mu\nu} 
- \tfrac{1}{2} {\cal D}_\mu \pi^a {\cal D}^\mu \pi_a \right) \longrightarrow 
S = S_0 - 
\tfrac{g}{2m^2} \int  d^Dx f^a_{\phantom{a}bc} F_a^{\mu\nu} {\cal D}_\mu \pi^b 
{\cal D}_\nu \pi^c\;,
\end{equation}
followed by the field redefinition \eqref{ExplFR}, keeping only 
cubic terms. 

Then one can formally follow the same procedure at each order in perturbation to find:\\
(1) The total field redefinition\footnote{This field redefinition is thus order-by-order invertible.} which maps from the full solution \eqref{fullsolution} to the solution with trivial gauge algebra;\\
(2) All the vertices at all orders which are obtained after doing 
the total field redefinition and which are by construction in the cohomology $H^0(\gamma)\,$, at ghost number zero.


\section{Massive gravity} \label{sec:massivegrav}

We now carry out the deformation analysis for a single massive spin-2 field 
on a maximally symmetric $D$-dimensional background. 
The components of the background metric are written as $g_{\mu\nu}\,$, 
while $g$ denotes its determinant and $\nabla_\mu$ the associated 
Levi-Civita covariant derivative. 
We introduce the parameter $\sigma$ that takes the value $1$ in 
the anti-de Sitter (AdS) background and the value $-1$ in the de Sitter 
(dS) geometry. 
The (A)dS radius will be denoted by $L$ and is related to the 
cosmological constant $\Lambda$ via $\Lambda = - \frac{(D-1)(D-2)}{2\,\sigma L^2}\,$. 
One can reach the Minkowski background by taking the limit 
$\Lambda \longrightarrow 0\,$, or equivalently, $L \longrightarrow \infty\,$. 
In our conventions, the commutator of background covariant 
derivatives acting on a (co)vector is 
$[\nabla_{\mu}, \nabla_{\nu}] V_\sigma = - \tfrac{2}{\sigma L^2} g_{\sigma[\mu} V_{\nu]}\,$.

The free theory is given by the massive Fierz--Pauli action in its Stueckelberg form,\footnote{Note that the physical graviton mass is related to the scale $m$ by $m^2_{\rm graviton}=2m^2$. Our choice of parametrization simplifies many of the equations that follow.}
\beq\bal
S_0&=\int d^Dx \Big[-\tfrac{1}{2}\,
\nabla^{\rho}h^{\mu\nu}\nabla_{\rho}h_{\mu\nu}+\nabla_{\rho}h^{\mu\nu}\nabla_{\mu}
h^{\rho}
_{\phantom{\rho}\nu}-\nabla_{\mu}h\nabla_{\nu}h^{\mu\nu}+\tfrac{1}{2}\,\nabla^{\mu}h\nabla_{\mu}h-\tfrac{1}{4}\,F^{\mu\nu}
F_{\mu\nu}\\
&\qquad\quad-\tfrac{1}{2}\,(\nabla\varphi)^2+2m\left(h\nabla_{\mu}B^{\mu}-h^{\mu\nu}\nabla_{\mu}B_{\nu}\right)+
\mu\,B^{\mu}\nabla_{\mu}\varphi - \left(\tfrac{D-1}{\sigma L^2} + m^2\right)h^{\mu\nu}h_{\mu\nu}\\
&\qquad\quad+\tfrac{1}{2}\left(\tfrac{D-1}{\sigma L^2}+ 2 m^2
\right)h^2-\mu m\,h\varphi+\tfrac{Dm^2}{D-2}\,\varphi^2-\tfrac{D-1}{\sigma L^2}\,B^{\mu}
B_{\mu}\Big]\sqrt{-g}\;,
\eal\eeq
with Stueckelberg fields $B_{\mu}$ and $\varphi\,$.
We use the notation $F_{\mu\nu}\coloneqq2\,\nabla_{[\mu}B_{\nu]}$ and introduce 
the following parameter $\mu$ with dimension of mass:
\beq
\mu\coloneqq\sqrt{\frac{2(D-1)}{D-2}}\sqrt{\frac{D-2}{\sigma L^2}+2 m^2}\;.
\eeq
In these conventions\footnote{As a result the partially massless point can only 
be obtained in the dS background. For conventions where the partially massless limit 
is possible with AdS as background but manifestly non-unitary (albeit real) 
Lagrangian, see \cite{Boulanger:2018shp}.} 
the fields are all canonically normalized and the action is manifestly 
unitary as far as the real mass parameter $m$ has a value such that 
the parameter $\mu$ is real. 
There are two gauge symmetries given by
\beq
\delta_{\xi}h_{\mu\nu}=2\nabla_{(\mu}\xi_{\nu)}\,,\qquad \delta_{\xi}B_{\mu}=-2m\,\xi_{\mu}\,,\qquad \delta_{\xi}\varphi=0\,,
\eeq
\beq
\delta_{\epsilon}h_{\mu\nu}=\tfrac{2\,m}{D-2}\,g_{\mu\nu}\epsilon\,,\qquad \delta_{\epsilon}B_{\mu}=\nabla_{\mu}\epsilon\,,
\qquad \delta_{\epsilon}\varphi=\mu\,\epsilon\,.
\eeq
The ghosts corresponding to these symmetries will be denoted by $C_{\mu}$ and $C$, respectively. The free part of the BV functional is then given by
\beq\bal
W_0&=S_0+\int\sqrt{-g}\Big[h^{*\mu\nu}\left(2\nabla_{(\mu}C_{\nu)}+\tfrac{2\,m}{D-2}\,g_{\mu\nu}C\right)
+B^{*\mu}\left(\nabla_{\mu}C-2\,mC_{\mu}\right)+\mu\,\varphi^{*}C\Big]\,.
\eal\eeq
The action of the differential $\gamma$ on the fields is
\beq\bal \label{eq:mg gammafields}
\gamma h_{\mu\nu}&=2\nabla_{(\mu}C_{\nu)}+\tfrac{2m}{D-2}\,g_{\mu\nu}C\;,
\quad\gamma B_{\mu}=\nabla_{\mu}C-2mC_{\mu}\;,
\quad\gamma \varphi=\mu\,C\,.
\eal\eeq
For the action of $\delta$ on the antifields, we have
\beq\bal
\delta h^{*\mu\nu}&=\Box h^{\mu\nu}-2\nabla_{\rho}\nabla^{(\mu}h^{\nu)\rho}+\nabla^{\mu}\nabla^{\nu}h+g^{\mu\nu}
\left(\nabla_{\rho}\nabla_{\sigma}h^{\rho\sigma}-\Box h\right)\\
&\quad-2\,m\left(\nabla^{(\mu}B^{\nu)}-g^{\mu\nu}\nabla_{\rho}B^{\rho}\right)-2 \left(\tfrac{D-1}{\sigma L^2} + m^2\right)h^{\mu\nu}\\
&\quad+\left(\tfrac{D-1}{\sigma L^2}+ 2 m^2
\right)g^{\mu\nu}h-\mu m\,g^{\mu\nu}\varphi\,,
\eal\eeq
\beq\bal
\delta B^{*\mu}&=\nabla_{\nu}F^{\nu\mu}+2m\left(\nabla_{\nu}h^{\mu\nu}-\nabla^{\mu}h\right)+\mu\,
\nabla^{\mu}\varphi-\tfrac{2(D-1)}{\sigma L^2}\,B^{\mu}\,,
\eal\eeq
\beq
\delta\varphi^{*}=\Box\varphi-\mu\left(\nabla_{\mu}B^{\mu}+mh\right)+\tfrac{2 Dm^2}{D-2}\,\varphi\,,
\eeq
while on the antighosts it is
\beq
\delta C^{*\mu}=-2\nabla_{\nu}h^{*\mu\nu}-2mB^{*\mu}\;,\quad
\delta C^{*}=\tfrac{2 m}{D-2}\,h^{*}-\nabla_{\mu}B^{*\mu}+\mu\,\varphi^{*}\;.
\eeq

\subsection{Cubic deformations}\label{cubicMassive}

We follow the same steps as in the case of massive Yang--Mills theory. 
Our guiding principle is to reproduce the cubic
vertex of the Einstein--Hilbert Lagrangian with cosmological constant, 
plus other terms with no more than two derivatives. This principle was also followed in the analyses of 
Zinoviev, see e.g.\ \cite{Zinoviev:2006im,Zinoviev:2011fv,Zinoviev:2014zka,Zinoviev:2016tua}.
It turns out that, for this to be the case, one must take the following
expression for the gauge algebra deforming term $a_2\,$, 
\beq \label{eq:mg a2}
a_2=-\kappa\, C^{*\mu}C^{\nu}\nabla_{\nu}C_{\mu}+g\,C^{*\mu}C_{\mu}C\,,
\eeq
with deformation parameters $\kappa$ and $g\,$. 
The first term is what leads to the Einstein--Hilbert cubic vertex 
\cite{Boulanger:2000rq}, with the constant $\kappa$ that coincides 
with $1/M_P^{(D-2)/2}\,$, $M_P$ being the Planck mass. 
The second term is needed in order to lift $a_2$ to a cubic vertex
$a_0$ with at most two derivatives. 

The two contractions in \eqref{eq:mg a2} 
can be lifted independently, that is they both satisfy the master 
equation at antifield number one, $\gamma a_1+\delta a_2+\td=0$. Defining
\beq \label{eq:mg gr and weyl}
a_2^{\rm (GR)}=C^{*\mu}C^{\nu}\nabla_{\nu}C_{\mu}\,,\qquad 
a_2^{\rm (extra)}=C^{*\mu}C_{\mu}C\,,
\eeq
we eventually obtain
\beq\bal
a_1^{\rm (GR)}&=-h^{*\mu\nu}\left(C^{\rho}\nabla_{\rho}h_{\mu\nu}+2\nabla_{(\mu}C^{\rho}h_{\nu)\rho}\right)-\tfrac{2m}
{D-2}\,h^{*\mu\nu}Ch_{\mu\nu}+\tfrac{2m}{D-2}\,h^{*}C^{\mu}B_{\mu}\\
&\quad+m\,B^{*\mu}C^{\nu}h_{\mu\nu}+\tfrac{1}{2}\,B^{*\mu}C^{\nu}F_{\mu\nu}-\tfrac{2 m^2}{\mu(D-2)}\,B^{*\mu}
C_{\mu}\varphi\,,
\eal\eeq
\beq
a_1^{\rm (extra)}=h^{*\mu\nu}Ch_{\mu\nu}-2h^{*\mu\nu}C_{\mu}B_{\nu}+\tfrac{2m}{\mu}\,B^{*\mu}C_{\mu}\varphi\,.
\eeq
Next we must also consider the homogeneous terms 
satisfying $\gamma\bar{a}_1+\td=0\,$. 
There are several candidates with at most one derivative 
in the gauge transformation laws. 
A priori, we could say that none of them are interesting since anyway they 
lead to trivial vertices, obtained from the free theory by a redefinition of the fields. 
Among them, there is a single $\bar{a}_1$ that we will consider for the reason that, when
added to $a_1 := -\kappa a_1^{\rm (GR)} + g a_1^{\rm (extra)}\,$ with a well-chosen 
coefficient, it gives a cubic vertex containing at most two 
derivatives in total. The other candidates do not have the same effect.
Explicitly, the $\bar{a}_1$ that enables one to obtain a two-derivative
cubic vertex is
\beq\bal \label{eq:mg a1bar}
\bar{a}_1&=\varphi^{*}C^{\mu}\left(\nabla_{\mu}\varphi-\mu\,
B_{\mu}\right)\;.
\eal\eeq
Thus, at this stage we have the antifield-1 solution
\beq \label{eq:mg a1sol}
a_1=-\kappa \,a_1^{\rm (GR)}+g\,a_1^{\rm (extra)}+\beta\,\bar{a}_1\,,
\eeq
with three free parameters $\kappa$, $g$ and $\beta$.

At the last step of the resolution of the master equation at first order 
in deformation, when solving $\gamma a_0+\delta a_1+\td=0\,$ for 
the vertex $a_0\,$, we find a unique solution for $a_0$ that 
requires the constants in \eqref{eq:mg a1sol} to take the following values:
\beq \label{eq:mg g and beta}
g=-\frac{m\kappa}{2}\,,\qquad \beta=\left[\frac{D+2}{4(D-1)}+\frac{D-4}{2 \mu^2 \,\sigma L^2}\right]\kappa\,.
\eeq
The resulting cubic vertex has a lengthy expression \eqref{fullvertex} that
we give in Appendix \ref{sec:app1}.
In the unitary gauge, it reduces to
\beq \label{eq:mg a0sol}
a_0\Big|_{B_{\mu}=0=\varphi}=\kappa \,\Lag_{\rm EH}^{(3)}+\kappa\, m^2\left(h^{\mu\nu}h_{\mu}^{\phantom{\mu}\rho}h_{\nu\rho}-
\tfrac{5}{4}\,hh^{\mu\nu}h_{\mu\nu}+\tfrac{1}{4}\,h^3\right)\;,
\eeq
with $\Lag_{\rm EH}^{(3)}$ the cubic part of the Einstein--Hilbert Lagrangian. The potential vertex proportional to $m^2$ 
coincides with the one found in \cite{Zinoviev:2006im,Zinoviev:2013hac}, 
and is a particular 
member of the dRGT class of massive gravity theories (see \cite{Hinterbichler:2011tt,deRham:2014zqa} for reviews). 
In fact, the complete solution \eqref{eq:mg a0sol} happens to be special, 
in that it is the unique massive graviton cubic interaction consistent 
with positivity constraints of eikonal scattering 
amplitudes and absence of superluminality \cite{Hinterbichler:2017qyt} (see also 
\cite{Cheung:2016yqr,Bonifacio:2016wcb,Bellazzini:2017fep,deRham:2017xox,Bonifacio:2017nnt,Bonifacio:2018vzv,deRham:2018qqo} 
for other 
studies of the $S$-matrix in massive gravity). It is also interesting that, 
in $D=4$, this vertex corresponds to the unique nonlinear action of a 
partially massless spin-2 
field \cite{Zinoviev:2006im,deRham:2012kf,Hassan:2012gz,deRham:2013wv}, 
although as is well known the theory happens to be obstructed at higher 
orders \cite{Deser:2013uy,Joung:2014aba,Garcia-Saenz:2014cwa} (more on this below).

This is not the end of the story, as we still have the option of adding homogeneous solutions $\bar{a}_0$, such that $
\gamma\bar{a}_0+\td=0$. Recall from the theorem of section \ref{subsec:stueck} that all such solutions are in fact of the Born--Infeld type, i.e.\ they satisfy $\gamma\bar{a}_0=0$. The goal is therefore to determine the cohomology of the differential $
\gamma$ in the fields. Interestingly, the answer turns out to be very simple:
\beq
H^0(\gamma)\cong\left\{f[H_{\mu\nu}]\right\}\,,
\eeq
where
\beq \label{eq:H tensor}
H_{\mu\nu}\coloneqq h_{\mu\nu}+\tfrac{1}{m}\,\nabla_{(\mu}B_{\nu)}
-\tfrac{1}{\mu m}\,\nabla_{\mu}\nabla_{\nu}
\varphi-\tfrac{2m}{\mu(D-2)}\,g_{\mu\nu}\varphi\,,
\eeq
is the unique Stueckelberg gauge invariant combination of the fields, 
as all other invariants can be built out of $H_{\mu\nu}$ 
and its derivatives. In particular the linearized Weyl tensor (which must 
clearly be invariant since the transformation of 
$h_{\mu\nu}$ is nothing but a linear diffeomorphism plus a Weyl rescaling) 
can be expressed in terms of appropriately projected second derivatives of $H_{\mu\nu}\,$.

The $\bar{a}_0$ vertices we seek are thus given by all the possible cubic combinations of the tensor $H_{\mu\nu}$ and 
derivatives thereof. 
At the lowest order in derivatives we can simply take contractions of powers of
$H_{\mu\nu}$ with no extra derivatives. 
But in fact there are more possibilities, since the tensors
\beq
\ytableausetup
 {mathmode, boxsize=0.2em}
H_{\mu\nu\rho}^{\ydiagram{2,1}}\coloneqq\nabla_{\mu}H_{\nu\rho}-\nabla_{\nu}H_{\mu\rho}\;,
\eeq
\beq
\ytableausetup
 {mathmode, boxsize=0.2em}
H_{\mu\nu\rho\sigma}^{\ydiagram{2,2}}\coloneqq
\nabla_{\mu}\nabla_{[\rho}H_{\sigma]\nu}-\nabla_{\nu}\nabla_{[\rho}H_{\sigma]\mu} + 
\nabla_{\rho}\nabla_{[\mu}H_{\nu]\sigma}-\nabla_{\sigma}\nabla_{[\mu}H_{\nu]\rho}\;,
\eeq
also contain no more than two derivatives and can be used to construct gauge invariant vertices. As the notation suggests, 
these correspond respectively to the ``hook'' and ``window'' Young projections of the first and second derivatives of 
$H_{\mu\nu}$. Now, generic cubic contractions of $H_{\mu\nu}$, $\ytableausetup{mathmode, boxsize=0.2em}H_{\mu\nu\rho}
^{\ydiagram{2,1}}$, and $\ytableausetup{mathmode, boxsize=0.2em}H_{\mu\nu\rho\sigma}^{\ydiagram{2,2}}$ will contain six 
derivatives, but there are three special combinations, namely
\beq \label{eq:mg drgt a0bar}
\bar{a}_0^{\rm(dRGT)}=\epsilon^{\mu_1\cdots\mu_3\rho_1\cdots\rho_{D-3}}\epsilon^{\nu_1\cdots\nu_3}
_{\phantom{\nu_1\cdots\nu_4}\rho_1\cdots\rho_{D-3}}H_{\mu_1\nu_1}H_{\mu_2\nu_2}H_{\mu_3\nu_3}\,,
\eeq
\beq \label{eq:mg pl1 a0bar}
\ytableausetup
 {mathmode, boxsize=0.2em}
\bar{a}_0^{\rm(PL_1)}=\epsilon^{\mu_1\cdots\mu_4\rho_1\cdots\rho_{D-4}}\epsilon^{\nu_1\cdots\nu_4}
_{\phantom{\nu_1\cdots\nu_4}\rho_1\cdots\rho_{D-4}}H_{\mu_1\nu_1\mu_2\nu_2}^{\ydiagram{2,2}}H_{\mu_3\nu_3}
H_{\mu_4\nu_4}\,,
\eeq
\beq \label{eq:mg pl2 a0bar}
\ytableausetup
 {mathmode, boxsize=0.2em}
\bar{a}_0^{\rm(PL_2)}=\epsilon^{\mu_1\cdots\mu_5\rho_1\cdots\rho_{D-5}}\epsilon^{\nu_1\cdots\nu_5}
_{\phantom{\nu_1\cdots\nu_5}\rho_1\cdots\rho_{D-5}}H_{\mu_1\nu_1\mu_2\nu_2}^{\ydiagram{2,2}}
H_{\mu_3\nu_3\mu_4\nu_4}^{\ydiagram{2,2}}H_{\mu_5\nu_5}\,,
\eeq
which actually contain only four derivatives (upon integration by parts).\footnote{Note that $\bar{a}_0^{\rm(PL_1)}$ and $\bar{a}_0^{\rm(PL_2)}$ may be alternatively written in terms of $\ytableausetup{mathmode, boxsize=0.2em}H_{\mu\nu\rho}
^{\ydiagram{2,1}}$ after integrating by parts.} When $\bar{a}_0^{\rm(dRGT)}$ is added to the solution 
of the deformation procedure, eq.\ \eqref{eq:mg a0sol}, it yields the full one-parameter cubic interactions of dRGT massive 
gravity. The terms $\bar{a}_0^{\rm(PL)}$ on the other hand correspond to the so-called pseudo-linear vertices analyzed in \cite{Hinterbichler:2013eza,Bonifacio:2018van}. Note however that $\bar{a}_0^{\rm(PL_2)}$ is only nontrivial in $D\geq5$ dimensions.

\subsection{Partially massless theory}
\label{sec.PM}

As mentioned before, there exists a unique cubic vertex for a partially massless (PM) spin-2 field in four dimensions, which 
moreover happens to belong to the dRGT class of massive gravity theories. Since we have recovered all the cubic 
interactions of dRGT theory via the deformation analysis, it is clear that the PM vertex should appear as a by-product. Indeed, 
we observed above that the solution in unitary gauge \eqref{eq:mg a0sol} of the descent equations precisely matches the cubic PM Lagrangian 
previously found via other methods \cite{deRham:2013wv}.

But this conclusion is a bit too quick since the calculations 
that led to Eq.\ \eqref{eq:mg a0sol} involved the use of the relation 
$C=\tfrac{1}{\mu}\gamma\varphi$, and now 
$\mu\coloneqq\sqrt{\tfrac{2(D-1)}{D-2}}\sqrt{\tfrac{D-2}{\sigma L^2}+2 m^2}=0$ 
because $\sigma=-1$ (dS background) and $2 L^2m^2=(D-2)$ at the PM point. 
Thus, for a PM spin-2, the ghost 
$C$ is now in the cohomology of $\gamma\,$, which turns out to significantly differ from the one associated with the massive set-up, and generates a standard (as opposed to Stueckelberg) gauge symmetry, as we 
of course expected. In practical terms this means that any step where we divided by $\mu$ is not allowed. A first implication is 
that the antifield-2 terms $a_2^{\rm (GR)}$ and $a_2^{\rm (extra)}$ in Eq.\ \eqref{eq:mg gr and weyl} can no longer be lifted 
independently, but instead we have
\beq \label{eq:mg a1pm}
a_1^{\rm (PM)}=-\kappa\left(a_1^{\rm (GR)} +\tfrac{m}{D-2}\,a_1^{\rm (extra)}\right)\,.
\eeq
The combination in parenthesis is finite at the PM point and moreover does not depend on $\varphi\,$, which is decoupled from the beginning. For that reason no $\bar{a}_1$ should be considered since the only one used in the massive case, Eq.\ \eqref{eq:mg a1bar}, contains $\varphi\,$.

Continuing with the last descent equation we find that 
$a_1^{\rm (PM)}$ can be lifted to a cubic vertex only if 
one sets
\beq
D=4\;.
\eeq
The expression for the vertex is given in \eqref{PMvertex}.
Actually this solution cannot directly be obtained from 
our results concerning the massive case and taking the PM 
limit \eqref{PMlimitvertex} in $D=4$ as some terms diverge and the scalar 
mode does not decouple from the other modes. Moreover, \eqref{eq:mg g and beta} would have given $\beta = \tfrac{\kappa}{2}$ 
while it should have been zero since \eqref{eq:mg a1bar} involves the scalar field. 
In order to see the PM case as a limit of the massive case, one should therefore 
treat the generic massive theory without \eqref{eq:mg a1bar}, which is in reality 
nothing else than doing a field redefinition of $\varphi$, since \eqref{eq:mg a1bar} 
is $\gamma$-exact, but the resulting vertices will have more than two derivatives. In fact all these higher-derivative terms contain the scalar field which decouples in the PM limit, and this is why one can find a vertex with no more than two derivatives in the PM case without adding \eqref{eq:mg a1bar}. We also expect other important features peculiar to a PM field to show up at the next order in the deformation analysis, since it is known 
that PM theory is obstructed at quartic order whereas massive gravity is not \cite{Deser:2013uy,Joung:2014aba,Garcia-Saenz:2014cwa}.

So far we have ignored the homogeneous solutions satisfying $\gamma\bar{a}_0+\td=0$. Here the analysis is quite different in 
PM relative to the generic case, since obviously the tensor $H_{\mu\nu}$ defined in \eqref{eq:H tensor} no longer makes 
sense. If we define instead
\beq
{\cal H}_{\mu\nu}\coloneqq h_{\mu\nu}+\tfrac{1}{m}\,\nabla_{(\mu}B_{\nu)}\,,
\eeq
we find that
\beq
{\cal F}_{\mu\nu\rho}\coloneqq\nabla_{\mu}{\cal H}_{\nu\rho}-\nabla_{\nu}{\cal H}_{\mu\rho}\,,
\eeq
is $\gamma$-closed and generates the cohomology of $\gamma$ in the fields; the tensor ${\cal F}_{\mu\nu\rho}$ is in fact 
nothing but the field strength of a PM graviton after one fixes unitary gauge. The possible Born--Infeld type vertices, $
\gamma\bar{a}_0^{(\rm BI)}=0$, are therefore given by all the possible contractions of ${\cal F}_{\mu\nu\rho}$ and its 
derivatives. Importantly, the dRGT and pseudo-linear terms in eqs.\ \eqref{eq:mg drgt a0bar}--\eqref{eq:mg pl2 a0bar} are not 
allowed in the PM case. Moreover, because the ghost $C$ does not correspond to a Stueckelberg symmetry anymore, the 
theorem of section \ref{sec:massivebv} does not apply and the possibility of constructing Chern--Simons type vertices for a 
PM spin-2 field is open. A complete analysis of the PM set-up in the BRST-BV formalism, including multiple fields (for which cubic vertices were already studied in \cite{Garcia-Saenz:2015mqi}) as well as the coupling to massless spin-2 particles, will be presented elsewhere.

\subsection{Triviality of the gauge algebra}
\label{sec 4.3}

The procedure described in subsection \ref{subsec:triviality} in the 
context of massive Yang--Mills can be applied in a very analogous way 
to massive gravity. The result will make it explicit that the cubic vertex 
\eqref{eq:mg a0sol} (see \eqref{fullvertex} for the full expression) 
can be written as a Born--Infeld type vertex followed by a field redefinition.

Writing $a_2$ as
\beq\bal
a_2&=-\kappa\, C^{*\mu}C^{\nu}\nabla_{\nu}C_{\mu}-\tfrac{m\kappa}{2}\,C^{*\mu}C_{\mu}C\\
&= \gamma \left( \tfrac{\kappa}{2}\, C^{*\mu} C^\nu (h_{\mu\nu} + \tfrac{1}{2m} F_{\mu\nu}) + \tfrac{m\kappa}{\sqrt{2}\,\mu} \left(\tfrac{D-4}{D-2}\right) C^{*\mu} C_\mu \varphi \right) = a_2^t\,,
\eal\eeq
allows us to read off the redefinition of the gauge parameter
\beq
\xi_\mu \longrightarrow \xi_\mu + \tfrac{\kappa}{2}\, (h_{\mu\nu} + \tfrac{1}{2m} F_{\mu\nu}) \xi^\nu + \tfrac{m\kappa}{\sqrt{2}\,\mu} \left(\tfrac{D-4}{D-2}\right) \varphi\, \xi_\mu 
\eeq
that trivializes the cubic deformation of the gauge algebra \eqref{eq:mg a2}.

The next step in the procedure is the resolution of the analogue of Eq.\ \eqref{diffa1gammaX}, which gives the information about the field redefinition. Here we have to solve for $c_1$ in
\beq \label{eq:mg c1 expression}
\delta \left( \tfrac{\kappa}{2}\, C^{*\mu} C^\nu (h_{\mu\nu} + \tfrac{1}{2m} F_{\mu\nu}) + \tfrac{m\kappa}{\sqrt{2}\,\mu} \left(\tfrac{D-4}{D-2}\right) C^{*\mu} C_\mu \varphi \right) - a_1 = \gamma c_1 + \td\,,
\eeq
where $a_1$ is the one given in \eqref{eq:mg a1sol} with the deformation parameters related through \eqref{eq:mg g and beta}. The simplest solution of this equation is rather lengthy and will be given in \eqref{c1massivegravity} in appendix \ref{sec:app1}. From the resulting $c_1$ we obtain the field redefinition \eqref{FRspin2trivial} that makes the gauge transformation trivial, that is equal to that of the free theory we started with.

Lastly the Born--Infeld vertex is given by
\beq
\bar{a}_0 = a_0 + \delta c_1\,,
\eeq
where the explicit expressions for $a_0$ and $c_1$ can be found in the 
appendix \ref{sec:app1}, respectively in \eqref{fullvertex} and
\eqref{c1massivegravity}. 
We emphasize again that the fact that this vertex is of the Born--Infeld 
type follows by construction, see Eq.\ \eqref{constrBI}. It is actually 
easy to construct $\bar{a}_0$ in its manifestly gauge invariant form, 
i.e.,\ in terms of $H_{\mu\nu}\,$. 
Indeed, any analytic function $f(h_{\mu\nu},B_{\mu},\varphi)$ which is 
gauge-invariant, $\gamma f = 0$, can be written as
\beq\label{fhBfequalsfH}
f(h_{\mu\nu},B_{\mu},\varphi) = f(h_{\mu\nu},B_\mu,\varphi)\Big|_{B_{\mu}=0=\varphi\,,\;h_{\mu\nu}\rightarrow H_{\mu\nu}}\,.
\eeq
This can be proven by observing the following:\\
(1) Any strictly gauge invariant function can be written as a function of 
$H_{\mu\nu}$ (and its derivatives):
\beq\label{ftildefH}
f(h_{\mu\nu},B_{\mu},\varphi) = \tilde{f}(H_{\mu\nu})\,;
\eeq
(2) Any analytic function of $h_{\mu\nu}\,$, $B_\mu$ and $\varphi$ can be split as a function of solely $h_{\mu\nu}$ plus a function of all the fields which vanishes in unitary gauge:
\beq
f(h_{\mu\nu},B_{\mu},\varphi) = f_1(h_{\mu\nu}) + f_2(h_{\mu\nu},B_{\mu},\varphi)\,,\qquad f_2(h_{\mu\nu},B_\mu,\varphi)\Big|_{B_{\mu}=0=\varphi} = 0\,.
\eeq
Combining these two equations one obtains $\tilde{f}(H_{\mu\nu}) = f_1(h_{\mu\nu}) + f_2(h_{\mu\nu},B_{\mu},\varphi)$, which implies, upon going to unitary gauge, $\tilde{f}(h_{\mu\nu}) = f_1(h_{\mu\nu})\,$. Using this in \eqref{ftildefH} proves the result \eqref{fhBfequalsfH}.

Thus $\bar{a}_0$ can be easily written in compact form as
\beq\bal
\bar{a}_0 &= (a_0 + \delta c_1)\Big|_{B_{\mu}=0=\varphi\,,\;h_{\mu\nu}\rightarrow H_{\mu\nu}}\\
&= \kappa \,\Lag_{\rm EH}^{(3)}(H_{\mu\nu})+\kappa\, m^2\left(H^{\mu\nu}H_{\mu}^{\phantom{\mu}\rho}H_{\nu\rho}-
\tfrac{5}{4}\,HH^{\mu\nu}H_{\mu\nu}+\tfrac{1}{4}\,H^3\right)\\
& \quad+ \tfrac{\kappa}{4} \Big[ \Box H^{\mu\nu}-2\nabla_{\rho}\nabla^{(\mu}H^{\nu)\rho}+\nabla^{\mu}\nabla^{\nu}H+g^{\mu\nu}
\left(\nabla_{\rho}\nabla_{\sigma}H^{\rho\sigma}-\Box H\right) \\
&\quad \qquad -2 \left(\tfrac{D-1}{\sigma L^2} + m^2\right)H^{\mu\nu} +\left(\tfrac{D-1}{\sigma L^2}+ 2 m^2
\right)g^{\mu\nu}H \Big] H_{\mu\lambda} H_{\nu}{}^{\lambda}\,. 
\eal\eeq

In conclusion, at the cubic level, the deformation obtained by deforming solely the action --- and not the gauge transformations --- by the addition of this $\bar{a}_0$ is equivalent to the deformation performed in section \ref{cubicMassive}, as they differ by a redefinition of the fields and gauge parameters of the theory.


\section{Discussion}

In this paper we have initiated the study of the consistent deformations of massive field theories using the BRST-BV formalism as a tool. Although such theories of course do not possess any gauge invariance in their usual parametrizations, it is well known that gauge symmetries can be straightforwardly introduced via the Stueckelberg procedure. Our goal in this work was two-fold: to understand the peculiarities of Stueckelberg gauge theories in the context of the BRST-BV formulation, and to see whether the method can be successfully applied to obtain consistent interaction vertices for massive fields.

Regarding the first question, we have unveiled an interesting picture of Stueckelberg models by showing that they always admit a choice of field variables in which the gauge algebra is manifestly Abelian, and with the action being constructed out of the gauge invariant building blocks of the initial free theory, what is called a Born--Infeld type model.

We illustrated these properties by studying the possible interaction vertices of massive Yang--Mills theory and of massive gravity on a maximally symmetric space. We showed that the BRST-BV formalism allows one to derive all the vertex structures, modulo some assumptions on the number of derivatives, that have been previously classified using other approaches. We also analyzed the special case in which the graviton is partially massless, and showed that the method allows us to recover, as a by-product, the known cubic vertex of PM gravity in four dimensions. We expect that the BRST-BV method will display its full potential with more complicated theories; we plan to tackle some of them in dedicated investigations, in particular the case of multiple massive and massless gravitons. It would also be interesting to gain further insight regarding the physically special spin-2 vertex, eq.\ \eqref{eq:mg a0sol}, that one gets if one were to ignore $\gamma$-closed solutions.

It is worth emphasizing that the interactions given by the BRST-BV deformation 
procedure are consistent in the sense that they preserve the number of gauge 
symmetries. On the other hand, in order for a deformation to maintain the number of 
degrees of freedom, one also ought to pay attention to Ostrogradski modes associated 
to equations of motion of order higher than two. The BRST-BV method is oblivious of 
this aspect, which therefore calls for a separate analysis, for instance a full 
Hamiltonian counting or a study of the decoupling limit theory. In this respect, 
however, massive theories in the Stueckelberg language are not at all different from 
standard gauge theories. The restriction on the number of derivatives appearing in 
the deformations is thus important for recovering ``healthy'' theories such as dRGT 
massive gravity.

\begin{acknowledgments}

We are grateful to A.\ Joyce, R.\ A.\ Rosen and especially Yu.\ Zinoviev for some very 
helpful conversations. Parts of our calculations were done using the Mathematica package {\it xTensor}. 
NB is Senior Research Associate of the F.R.S.-FNRS (Belgium) and his work 
was supported in part by a FNRS PDR grant number T.1025.14.
LT is Research Fellow of the F.R.S.-FNRS. They both want to thank IAP for 
kind hospitality. 
CD and SGS are supported by the European Research Council under the European Community's Seventh Framework Programme (FP7/2007-2013 Grant Agreement no.\ 307934, NIRG project). SGS also thanks UMONS for generous hospitality.

\end{acknowledgments}


\appendix

\section{Full cubic vertex of massive gravity} \label{sec:app1}

The full cubic vertex arising in the deformation of linearized massive gravity in the Stueckelberg formulation, and which reduces to \eqref{eq:mg a0sol} in the unitary gauge, is given by
\beq\bal
\label{fullvertex}
a_0 &=\kappa \,\Lag_{\rm EH}^{(3)}+\kappa\, m^2\left(h^{\mu\nu}h_{\mu}^{\phantom{\mu}\rho}h_{\nu\rho}-
\tfrac{5}{4}\,hh^{\mu\nu}h_{\mu\nu}+\tfrac{1}{4}\,h^3\right) + \tfrac{3m \kappa}{2} A^\sigma h^{\mu\nu} \nabla_\sigma h_{\mu\nu}\\
& \quad - m \kappa A^\mu h^{\nu\sigma} \nabla_\sigma h_{\mu\nu} - 2m \kappa A^\mu h_{\mu\nu} \nabla_\sigma h^{\nu\sigma} + \tfrac{m \kappa}{2} A_\mu h \nabla_\nu h^{\mu\nu} + \tfrac{3m \kappa}{2} A_\mu h^{\mu\nu} \nabla_\nu h \\
& \quad - \tfrac{m\kappa}{2} A_{\mu} h \nabla^\mu h - \tfrac{\kappa}{16} h F_{\mu\nu} F^{\mu\nu} - \tfrac{\kappa}{4} h_{\mu\nu} F^{\mu\sigma} F_\sigma{}^\mu + \tfrac{\kappa \tilde{\mu}^2}{2\mu^2} h^{\mu\nu} \nabla_\mu \varphi \nabla_\nu \varphi - \tfrac{\kappa \tilde{\mu}^2}{4 \mu^2} h \nabla_\mu \varphi \nabla^\mu \varphi\\
& \quad + \tfrac{m \kappa}{8 \mu} \left( \tfrac{D-4}{D-2} \right) \varphi F_{\mu\nu} F^{\mu\nu} - \tfrac{\kappa \tilde{\mu}^2}{\mu} h^{\mu\nu} A_\mu \nabla_\nu \varphi + \tfrac{\kappa \tilde{\mu}^2}{2 \mu} h A^\mu \nabla_\mu \varphi - \tfrac{m \kappa \tilde{\mu}^2}{\mu^2} A^\mu \varphi \nabla_\mu \varphi\\
& \quad + \tfrac{m \kappa \tilde{\mu}^2}{2 \mu^3} \varphi \nabla_\mu \varphi \nabla^\mu \varphi + \tfrac{m \kappa}{2 \mu} \left( \tfrac{2(D-1)}{\sigma L^2} + \tfrac{3 D m^2}{D-2} \right) \varphi h_{\mu\nu} h^{\mu\nu} - \tfrac{\kappa m \tilde{\mu}^2}{2\mu} \varphi h^2\\
& \quad + \tfrac{\kappa}{2} \left( \tfrac{D-1}{\sigma L^2} - m^2 \right) h_{\mu\nu} A^\mu A^\nu - \tfrac{\kappa}{4} \left( \tfrac{D-1}{\sigma L^2} + m^2 \right) h A_\mu A^\mu + \tfrac{m \kappa \tilde{\mu}^2}{2 \mu} \varphi A_\mu A^\mu \\
& \quad + \tfrac{m^2 \kappa}{2 \mu^2} \left( \tfrac{D-1}{D-2} \right) \left( \tfrac{D+2}{\sigma L^2} + \tfrac{12 m^2}{D-2} \right) h \varphi^2 - \tfrac{m^3 \kappa}{3 \mu^3} \tfrac{D(D-1)}{(D-2)^2} \left( \tfrac{D+2}{\sigma L^2} + \tfrac{12 m^2}{D-2} \right) \varphi^3\;,
\eal\eeq
where the Einstein--Hilbert cubic piece is written as
\beq\bal
\Lag_{\rm EH}^{(3)} &= \tfrac{1}{2} h^{\mu\nu} \nabla_\mu h^{\rho\sigma} \nabla_\nu h_{\rho\sigma}-\tfrac{1}{2} h^{\mu\nu} \nabla_\mu h \nabla_\nu h + 2 h^{\mu\nu} \nabla_\rho h_\mu{}^\rho \nabla_\nu h -  h^{\mu\nu} \nabla_\rho h \nabla^\rho h_{\mu\nu}\\
& \quad + \tfrac{1}{4} h \nabla^\rho h \nabla_\rho h - 2 h^{\mu\nu} \nabla_\rho h_\mu{}^\rho \nabla_\sigma h_\nu{}^\sigma - 2 h^{\mu\nu}
   \nabla_\sigma h_\rho{}^\sigma \nabla_\nu h_\mu{}^\rho + h \nabla_\nu h^{\nu\rho} \nabla_\sigma h_\rho{}^\sigma\\
   & \quad + h^{\mu\nu} \nabla_\sigma h_\nu{}^\sigma \nabla^\sigma h_{\mu\nu} - \tfrac{1}{2} h \nabla^\sigma h \nabla_\sigma h_\rho{}^\sigma + h^{\mu\nu} \nabla^\sigma h_\mu{}^\rho \nabla_\rho h_{\nu\sigma} + h^{\mu\nu} \nabla^\sigma h_\mu{}^\rho \nabla_\sigma h_{\nu\rho}\\
   & \quad - \tfrac{1}{2} h \nabla_\sigma h_{\nu\sigma} \nabla^\sigma h^{\nu\rho} - \tfrac{1}{4} h \nabla_\sigma h_{\nu\rho} \nabla^\sigma h^{\nu\rho} - \tfrac{2 (D + 2)}{3 \sigma L^2} h^{\mu\nu}h_{\mu}^{\phantom{\mu}\rho}h_{\nu\rho} + \tfrac{3}{\sigma L^2}  h h_{\mu\nu} h^{\mu\nu} +\tfrac{(D -7)}{6 \sigma L^2} h^3
\eal\eeq
and with the definition
\beq
\tilde{\mu} \coloneqq \sqrt{\frac{D-1}{\sigma L^2} + \left( \frac{D+2}{D-2} \right)m^2}
\eeq
for the parameter $\tilde{\mu}$ which has dimension of mass.

In the $dS_4$ background, the PM limit of this vertex is not finite but can be written as
\beq\bal
\label{PMlimitvertex}
a_0^{\rm (PM\,limit)} &= \kappa \,\Lag_{\rm EH}^{(3)}+ \lim\limits_{m\rightarrow\tfrac{1}{L}} \Bigg[ \kappa\, m^2\left(h^{\mu\nu}h_{\mu}^{\phantom{\mu}\rho}h_{\nu\rho}-
\tfrac{5}{4}\,hh^{\mu\nu}h_{\mu\nu}+\tfrac{1}{4}\,h^3\right) + \tfrac{3m \kappa}{2} A^\sigma h^{\mu\nu} \nabla_\sigma h_{\mu\nu}\\
& \quad - m \kappa A^\mu h^{\nu\sigma} \nabla_\sigma h_{\mu\nu} - 2m \kappa A^\mu h_{\mu\nu} \nabla_\sigma h^{\nu\sigma} + \tfrac{m \kappa}{2} A_\mu h \nabla_\nu h^{\mu\nu} + \tfrac{3m \kappa}{2} A_\mu h^{\mu\nu} \nabla_\nu h \\
& \quad - \tfrac{m\kappa}{2} A_{\mu} h \nabla^\mu h - \tfrac{\kappa}{16} h F_{\mu\nu} F^{\mu\nu} - \tfrac{\kappa}{4} h_{\mu\nu} F^{\mu\sigma} F_\sigma{}^\mu + \tfrac{\kappa}{4} h^{\mu\nu} \nabla_\mu \varphi \nabla_\nu \varphi \\
& \quad - \tfrac{\kappa}{8} h \nabla_\mu \varphi \nabla^\mu \varphi - \tfrac{m \kappa}{2} A^\mu \varphi \nabla_\mu \varphi -2 m^2 \kappa\, h_{\mu\nu} A^\mu A^\nu + \tfrac{m^2 \kappa}{2} h A_\mu A^\mu + \tfrac{3 m^2 \kappa}{4} h \varphi^2 \\
& \quad + \tfrac{m \kappa}{\sqrt{6}\sqrt{m^2-\tfrac{1}{L^2}}} \left( \tfrac{1}{4}\, \varphi \nabla_\mu \varphi \nabla^\mu \varphi - m^2\, \varphi^3 \right)\Bigg]\;.
\eal\eeq

Instead of doing the PM limit of the result \eqref{fullvertex}, one can follow the deformation procedure as in section \ref{sec.PM} to eventually obtain
\beq\bal
\label{PMvertex}
a_0^{\rm (PM)} &= \kappa \,\Lag_{\rm EH}^{(3)}+\kappa\, m^2\left(h^{\mu\nu}h_{\mu}^{\phantom{\mu}\rho}h_{\nu\rho}-
\tfrac{5}{4}\,hh^{\mu\nu}h_{\mu\nu}+\tfrac{1}{4}\,h^3\right) + \tfrac{3m \kappa}{2} A^\sigma h^{\mu\nu} \nabla_\sigma h_{\mu\nu}\\
& \quad - m \kappa A^\mu h^{\nu\sigma} \nabla_\sigma h_{\mu\nu} - 2m \kappa A^\mu h_{\mu\nu} \nabla_\sigma h^{\nu\sigma} + \tfrac{m \kappa}{2} A_\mu h \nabla_\nu h^{\mu\nu} + \tfrac{3m \kappa}{2} A_\mu h^{\mu\nu} \nabla_\nu h \\
& \quad - \tfrac{m\kappa}{2} A_{\mu} h \nabla^\mu h - \tfrac{\kappa}{16} h F_{\mu\nu} F^{\mu\nu} - \tfrac{\kappa}{4} h_{\mu\nu} F^{\mu\sigma} F_\sigma{}^\mu -2 m^2 \kappa\, h_{\mu\nu} A^\mu A^\nu + \tfrac{m^2 \kappa}{2} h A_\mu A^\mu\,,
\eal\eeq
where $m=\tfrac{1}{L}\,$. The above expression exactly is \eqref{PMlimitvertex} 
with the scalar field $\varphi$ put to zero before taking the limit.

In section \ref{sec 4.3} we calculated the field redefinition mapping the algebra-deforming vertices to the Born--Infeld-like deformation. This field redefinition is encoded in the object $c_1$ (see eq.\ \eqref{eq:mg c1 expression}) which is written in full as
\beq\bal \label{c1massivegravity}
c_1 &= \kappa h^{*\mu\nu} \Big[ \tfrac{1}{4} h_{\mu\sigma} h_\nu{}^\sigma -\tfrac{1}{4 m}h_{\mu}{}^\rho F_{\nu\rho} +\tfrac{1}{2 m}A^\rho
   \nabla_\nu h_{\mu\rho} -\tfrac{1}{2 m}A^\rho \nabla_\rho h_{\mu\nu} -\tfrac{1}{2 \mu  m}\nabla^\rho \varphi \nabla_\nu h_{\mu\rho}\\
   & \qquad \qquad + \tfrac{1}{2 \mu  m} \nabla^\rho \varphi \nabla_\rho h_{\mu\nu} - \tfrac{1}{4 \mu  m^2} \nabla^\rho \varphi \nabla_\nu F_{\mu\rho} + \tfrac{1}{2 \mu m^2\, \sigma L^2} g_{\mu\nu} A^\rho \nabla_\rho \varphi \\
   & \qquad \qquad - \tfrac{1}{2 \mu m^2} \left( \tfrac{1}{\sigma L^2} - \left( \tfrac{D-4}{D-2} \right) m^2 \right) A_\mu \nabla_\nu \varphi + \tfrac{1}{16 m^2} F_{\mu\rho} F_\nu{}^\rho + \tfrac{1}{4 m^2} A^\rho \nabla_\mu F_{\nu\rho} \\
   & \qquad \qquad + \tfrac{1}{4 \mu^2 m^2} \left( \tfrac{1}{\sigma L^2} - \left( \tfrac{D-6}{D-2} \right) m^2 \right) \nabla_\mu \varphi \nabla_\nu \varphi - \tfrac{1}{8 (D-1) m^2}g_{\mu\nu}
   \nabla_\rho \varphi \nabla^\rho \varphi - \tfrac{(D -4) m
   }{2 (D -2) \mu } \varphi h_{\mu\nu} \\
   & \qquad \qquad + \tfrac{1}{4 m^2} \left( \tfrac{1}{\sigma L^2}-m^2 \right) A_\mu A_\nu - \tfrac{1}{4 m^2} \left(\tfrac{1}{\sigma L^2}-\tfrac{2 m^2}{D -2}\right) g_{\mu\nu} A_\rho A^\rho + \tfrac{(D -4) m^2}{2 (D -2)^2 \mu ^2} g_{\mu\nu} \varphi^2 \Big]\;.
\eal\eeq
This translates into the field redefinition
\beq
\label{FRspin2trivial}
h_{\mu\nu} \longrightarrow h_{\mu\nu} + \tfrac{\kappa}{4}\, h_{\mu\sigma} h_\nu{}^\sigma + \dots
\eeq
where the dots refer to terms proportional to the spin-1 and spin-0 modes and vanish in the unitary gauge.


\bibliographystyle{apsrev4-1}
\bibliographystyle{utphys}

\bibliography{Biblio}



\end{document}